\documentclass[12pt]{iopart}
\usepackage{scalerel}
\usepackage{tikz}
\usepackage{multirow}
\usepackage{graphicx}
\usepackage{subcaption}

\usetikzlibrary{svg.path}

\definecolor{orcidlogocol}{HTML}{A6CE39}
\tikzset{
  orcidlogo/.pic={
    \fill[orcidlogocol] svg{M256,128c0,70.7-57.3,128-128,128C57.3,256,0,198.7,0,128C0,57.3,57.3,0,128,0C198.7,0,256,57.3,256,128z};
    \fill[white] svg{M86.3,186.2H70.9V79.1h15.4v48.4V186.2z}
                 svg{M108.9,79.1h41.6c39.6,0,57,28.3,57,53.6c0,27.5-21.5,53.6-56.8,53.6h-41.8V79.1z M124.3,172.4h24.5c34.9,0,42.9-26.5,42.9-39.7c0-21.5-13.7-39.7-43.7-39.7h-23.7V172.4z}
                 svg{M88.7,56.8c0,5.5-4.5,10.1-10.1,10.1c-5.6,0-10.1-4.6-10.1-10.1c0-5.6,4.5-10.1,10.1-10.1C84.2,46.7,88.7,51.3,88.7,56.8z};
  }
}

\newcommand\orcidicon[1]{\href{https://orcid.org/#1}{\mbox{\scalerel*{
\begin{tikzpicture}[yscale=-1,transform shape]
\pic{orcidlogo};
\end{tikzpicture}
}{0}}}}

\usepackage{graphicx}
\usepackage[utf8]{inputenc} 
\usepackage[T1]{fontenc}    
\usepackage{xcolor}
\definecolor{darkblue}{RGB}{46,48,147}
\usepackage[colorlinks=true,
            linkcolor=darkblue,
            urlcolor=darkblue,
            citecolor=darkblue]{hyperref}
\usepackage{url}            
\usepackage{booktabs}       
\usepackage{amsfonts}       
\usepackage{nicefrac}       
\usepackage{microtype}      
\usepackage{multirow}
\usepackage{xspace}
\usepackage{amssymb}
\usepackage{todonotes}
\usepackage{cite}
\usepackage{lineno}

\graphicspath{{figures/}}

\begin{document}

\title{GWAK: Gravitational-Wave Anomalous Knowledge with Recurrent Autoencoders}

\author{Ryan Raikman$^{1,5*}$, Eric A. Moreno$^2$, Ekaterina Govorkova$^2$, Ethan J Marx$^{1,2}$, Alec Gunny$^{1,2}$, William Benoit$^3$, Deep Chatterjee$^{1,2}$, Rafia Omer$^3$, Muhammed Saleem$^3$, Dylan S Rankin$^4$, Michael W Coughlin$^3$, Philip C Harris$^2$, Erik Katsavounidis$^{1,2}$}

\address{
$^1$MIT LIGO Laboratory, USA, \newline $^2$Massachusetts Institute of Technology, USA,\newline $^3$University of Minnesota, USA, \newline $^4$University of Pennsylvania, USA, \newline $^5$Carnegie Mellon University, USA}
\ead{$^*$rraikman@mit.edu}
\vspace{10pt}
\begin{indented}
\item[]July 2023
\end{indented}

\begin{abstract}
Matched-filtering detection techniques for gravitational-wave (GW) signals in ground-based interferometers rely on having well-modeled templates of the GW emission. 
Such techniques have been traditionally used in searches for compact binary coalescences (CBCs), and have been employed in all known GW detections so far.
However, interesting science cases aside from compact mergers do not yet have accurate enough modeling to make matched filtering possible, including core-collapse supernovae and sources where stochasticity may be involved.
Therefore the development of techniques to identify sources of these types is of significant interest.
In this paper, we present a method of anomaly detection based on deep recurrent autoencoders to enhance the search region to unmodeled transients.
We use a semi-supervised strategy that we name \textit{``Gravitational Wave Anomalous Knowledge''} (GWAK).
While the semi-supervised nature of the problem comes with a cost in terms of accuracy as compared to supervised techniques, there is a qualitative advantage in generalizing experimental sensitivity beyond pre-computed signal templates. 
We construct a low-dimensional embedded space using the GWAK method, capturing the physical signatures of distinct signals on each axis of the space.
By introducing signal priors that capture some of the salient features of GW signals, we allow for the recovery of sensitivity even when an unmodeled anomaly is encountered.
We show that regions of the GWAK space can identify CBCs, detector glitches and also a variety of unmodeled astrophysical sources.
\end{abstract}

\vspace{2pc}
\maketitle

\section{Introduction}
\label{sec:intro}

Since the original observation of gravitational waves (GW)~\cite{abbott2016observation} by Advanced LIGO~\cite{Aasi_2015} and Advanced VIRGO~\cite{Acernese_2015}, and with the recent introduction of KAGRA~\cite{KAGRA:2020tym}, more than 90 gravitational-wave events~\cite{PhysRevX.9.031040, LIGOScientific:2020ibl, LIGOScientific:2021djp} have been catalogued to date, fundamentally transforming our way of observing the universe.
While all of the detected signals thus far correspond to the coalescence of binary black hole (BBH), binary neutron star (BNS), or black hole - neutron star (BHNS) mergers~\cite{abbott2016gw151226,abbott2017gw170104,abbott2017gw170814,abbott2017gw170817,abbott2019gwtc}, there is growing interest in the detection of unmodeled signals, which are not described by any known theoretical waveforms, computationally prohibitive to simulate, or are stochastic in nature.
These signals may originate from sources such as core-collapse supernovae (CCSN)~\cite{Abdikamalov_2021}, as well as exotic sources such as cosmic strings~\cite{Hindmarsh_1995, Damour_2000}, axion stars~\cite{braaten2018axion}, neutron star glitches, or primordial black holes~\cite{franciolini2021primordial,Eroshenko_2018}. Detection of such sources could lead to new insights into the fundamental physics of the universe.

Typical methods for detecting GWs rely on matched filtering techniques~\cite{Allen:2005fk}, which compare the observed data to a known signal template. 
These methods require precise knowledge of the signal prior, such as the waveform and the parameters of the source, in order to detect the signal.
While matched filtering is a well-established and powerful method for detecting gravitational waves, it has certain limitations. 
For example, matched filtering is sensitive only to signals that match the known templates, and may miss signals with different waveforms or parameters.
This makes matched filtering unsuitable for unmodeled signal searches.

The GW community has developed several unmodeled approaches that do not rely on a specific waveform model. Some of these currently used by the International Gravitational-wave Network (IGWN) include cWB~\cite{PhysRevD.93.042004,PhysRevD.72.122002}, which searches for and reconstructs GW transient signals without relying on a specific waveform model. 
Another framework, oLIB~\cite{PhysRevD.95.104046} uses the Q transform to decompose GW strain data into several time-frequency planes of constant quality factors. The pipeline flags data segments containing excess power and searches for clusters of these segments to identify possible GW candidate events. MLy (read as ``Emily'')~\cite{skliris2022realtime} is a machine-learning-based search for generic sub-second-duration transient GW signals in the 20 to 500\,Hz frequency band; MLy works by utilizing convolutional neural networks (CNNs) trained to recognize signals that are simultaneous and coherent between detectors.

In this paper, we explore the use of a method introduced in Ref.~\cite{quak} deployed within the High Energy Physics community for the development of an anomaly detection pipeline for data collected by GW observatories. We introduce \textit{``Gravitational Wave Anomalous Knowledge''}~(GWAK), a strategy for anomaly search that combines deep learning~(DL) techniques with prior information on potential signals to improve the sensitivity of detection.
The GWAK algorithm is based on the intuition that unknown transient sources should loosely resemble known signals, as well as be coherent between present GW detectors.
We apply the GWAK method to GW datasets, by introducing signal priors that capture some of the salient features of GW signatures, allowing for the recovery of sensitivity even when the observed signal mismatches the known priors.

Because GWAK does not rely on precise prior knowledge of the signal and can detect signals with unknown waveforms or parameters by matching incoming data streams with salient features (cross-correlation, oscillations, etc.) that are generic to broad types of GWs, GWAK is more robust and powerful for the detection of unknown signals. 
As such, it can be used as a complementary approach to methods such as matched filtering for detecting transient GWs, and it has the potential to improve the sensitivity of GW detection systems to sources of this type.

This paper is organized as follows: in Section \ref{sec:relatedWorks}, we provide a brief review of deep learning in GW detection and previous works in machine learning anomaly detection. 
In Section \ref{sec:data}, we describe the data used for this study. 
In Section \ref{sec:GWAK}, we present the GWAK method for constructing embedded spaces for anomalous searches and the autoencoder and transformer architectures used to build these embedded spaces. 
In Section \ref{sec:results}, we discuss the performance of the GWAK method on real GW data. 
Conclusions and next steps are provided in Section \ref{sec:conclusions}.

The code used to analyze data and generate results and plots can be found at~\footnote{\url{https://github.com/ML4GW/gwak}}.

\section{Related work}
\label{sec:relatedWorks}
Deep learning approaches for GW detection are well explored~\cite{Baker:2014eba,George:2016hay, Kapadia:2017fhb,George:2017pmj,Gabbard:2017lja,Miller:2019jtp,Jadhav:2020oyt,Huerta:2020xyq, 9956104, Chatterjee_2021, beveridge2023detection}. 
However, these methods typically rely on supervised learning techniques, which provide competitive efficiency by exploiting neural network nonlinearity and the information provided by ground-truth labels.
By construction, these methods rely on a realistic simulation of the signal generated by a specific kind of source, which is assumed upfront.
With supervised DL approaches, there is no guarantee of generalizability to an out-of-training event, what we refer to as ``anomalies.'' 

Autoencoders are commonly implemented for a variety of GW applications. 
They can be used for non-linear subtraction of noise from incoming time-series of GW strain \cite{Ormiston:2020ele, Bacon:2022lsm, Saleem:2023hcm} in real-time \cite{Gunny:2021gne}. 
Additionally, autoencoders are used in generative models, speeding up the computation of GW waveforms relative to very computationally expensive numerical relativity simulations \cite{PhysRevD.103.124051}. 
Finally, due to the transient nature of single-detector artifacts, ``glitches'', autoencoders can be used for glitch classification in an unsupervised manner~\cite{Sankarapandian:2021qun}.

There have also been explorations into unsupervised GW detection to enhance detection capabilities beyond signal templates and simulation. 
Initial studies with a Convolutional Neural Network~(CNN) based autoencoder \cite{Morawski:2021kxv} and Long-Short Term Memory~(LSTM) based autoencoder~\cite{eric_moreno_2021_5772814} show that unsupervised detection is possible.
Both studies rely on learning the typical features of the background.
Once a model is trained, it is then used to evaluate the similarity between background priors and new unseen data. 
To do this, the algorithms use reconstruction loss, computed by comparing the original signal and the signal outputted by the model trained on the background prior, as a detection statistic.
By comparing the reconstruction error of new data with a threshold corresponding to an allowed false alarm rate~(FAR), data points that deviate enough from the normal pattern are identified as anomalous.
This relies on the autoencoder's inability to reconstruct any potential signal that deviates from the background, or generally the prior on which the model was trained, triggering a high reconstruction loss.
This approach has been shown to effectively detect out-of-training anomalies in GW datasets and has the potential to improve the performance of anomaly detection systems. 
However, for a specific signal, an algorithm trained with an unsupervised procedure on unlabeled data is typically less accurate than a supervised classifier trained on labeled data. 

We build upon this approach by training several autoencoders; each with a different signal prior to enhance signal sensitivity.
The signals that were chosen to be used in this paper are described in Sec.~\ref{sec:data}.
Unlike an earlier study~\cite{eric_moreno_2021_5772814} which used simulated Gaussian background as a proof of concept, our method is trained with real background data. This is significantly more challenging than just simulated Gaussian noise, and therefore no direct comparison of the results presented here to the previous ones can be made.

Although the autoencoder architecture can implicitly learn detector correlation, a related method explicitly leverages the correlation across detectors \cite{Skliris:2020qax}. 
The method trains detector correlation with a white-noise burst (WNB)~\cite{Sutton_2010, PhysRevD.95.042003} signal prior, which should be harder to detect than any other type of signal given its lack of a distinctive morphology.
Similarly to \cite{Skliris:2020qax}, we choose not to rely on our autoencoders to learn the correlation between the two detector sites. 
Instead, we directly compute and include the correlation in our final metric that is used to find signal events as another axis in the GWAK embedded space. As such, the present version of our method is applicable to the case of aligned GW detectors. However, more off-plane GW detectors are being proposed, and a future area of work is generalizing the method to an arbitrary detector network.

\section{GWAK algorithm}
\label{sec:GWAK}
GWAK~(reads: \textit{guac}) builds on the concept of semi-supervision, pulling on concepts from both supervised and unsupervised learning. 
The semi-supervised method is manifested by using simulated signals as approximations for anomalous-unmodeled signals in the GWAK embedded space.
We use five classes of datasets that can help us to build an informative space to search for these unmodeled signals.
A separate unsupervised autoencoder network is then trained on each class of samples separately, resulting in a low-dimensional ``GWAK'' space consisting of the coherence metrics between the autoencoder inputs and outputs for each autoencoder, which is then used to search for anomalous signals.
This approach is particularly useful for detecting new physics phenomena, where the signal prior is unknown but the simulation of some potential signal pattern, like BBH and sine-Gaussians, is available. The GWAK method results in classes of anomalous signals inhabiting different regions of the continuous GWAK space, each being reconstructed differently by the five autoencoders. Searches can then be performed on the lower-dimensional embedded space in the regions that anomalies are expected to inhabit. 

\begin{figure}[htb]
\centering
    \includegraphics[width=0.44\textwidth]{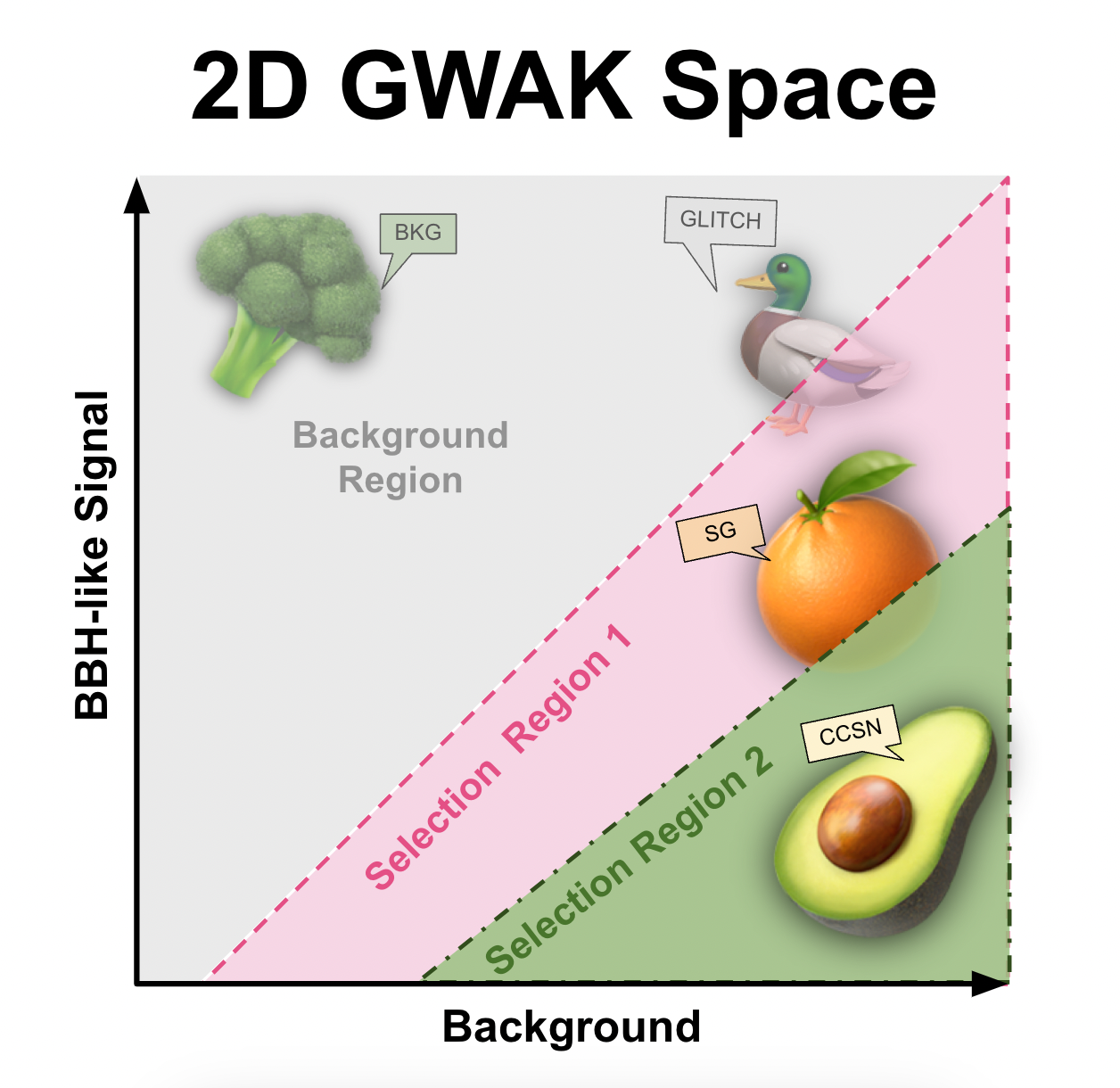}    
    \includegraphics[width=0.55\textwidth]{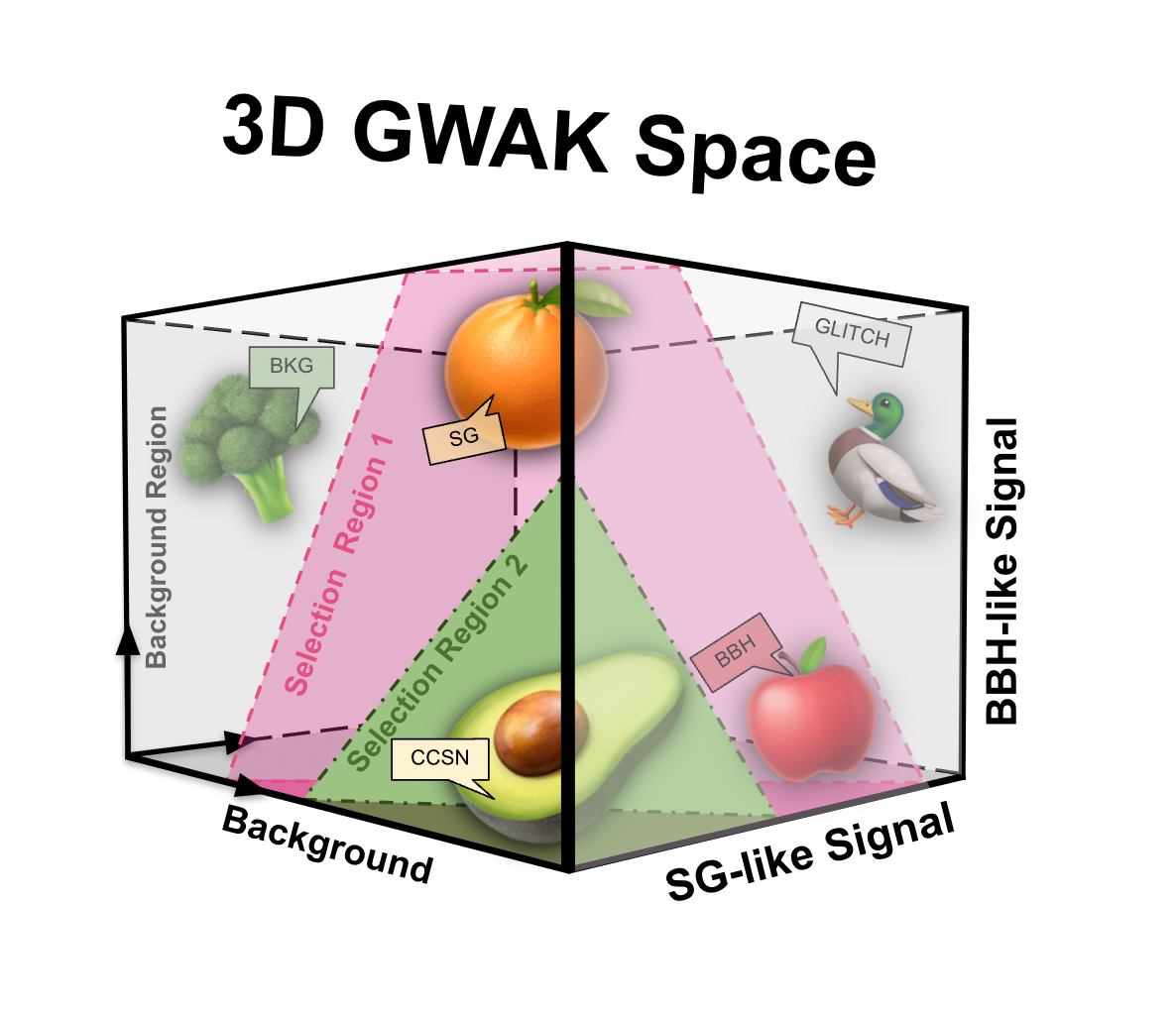}
\caption{The 2D and 3D schematic depicting the main idea behind the GWAK algorithm. For plotting purposes we only show one background and two signal axes, while in this paper we are actually using 5 (2 for background and 3 for signals), and the main idea remains the same. All the fruits (orange, avocado and apple) are ``signals'' that we would like to select, broccoli represents background that we would like to suppress and the duck represents true anomalies, such as detector malfunctioning, and glitches. Shaded pink and green region depicts selection regions that would correspond to different FARs. For higher FAR~(pink shading) more signal-like anomalies would be picked up (orange, avocado and apple), while for a selection that corresponds to a lower FAR (green shading), only ``avocado signal'' would be picked up.}
\label{fig:3d_gwak}
\end{figure}

\subsection{Network architectures}
\label{sec:AE}
One of the main advantages of LSTM autoencoders is their ability to handle sequential data with temporal dependencies. 
This makes them suitable for anomaly detection in time-series data, such as GWs, speech signals, and sensor data. 
The LSTM autoencoder consists of an encoder and a decoder, where the encoder maps the input sequence to a fixed-length vector, and the decoder maps the vector back to the original sequence. 
We use a similar architecture to that optimized in~\cite{eric_moreno_2021_5772814}.

For the signal classes, we used the LSTM autoencoder as described above, with the bottleneck size of 4, 8 and 8 for binary black hole (BBH), low and high frequency sine-gaussian (SG) signals, respectively. 
For the background classes, we used a fully connected dense model. This was done as the signal classes have temporal behavior to exploit, whereas glitches have smaller-scale, localized features.
The total number of trainable parameters for the BBH LSTM AE is 510324, for SG 64--512\,Hz and SG 512--1024\,Hz are 511672, for background AE 243352 and for glitches AE is 241302.

\begin{figure}[htb]
\centering
    \includegraphics[width=1\textwidth]{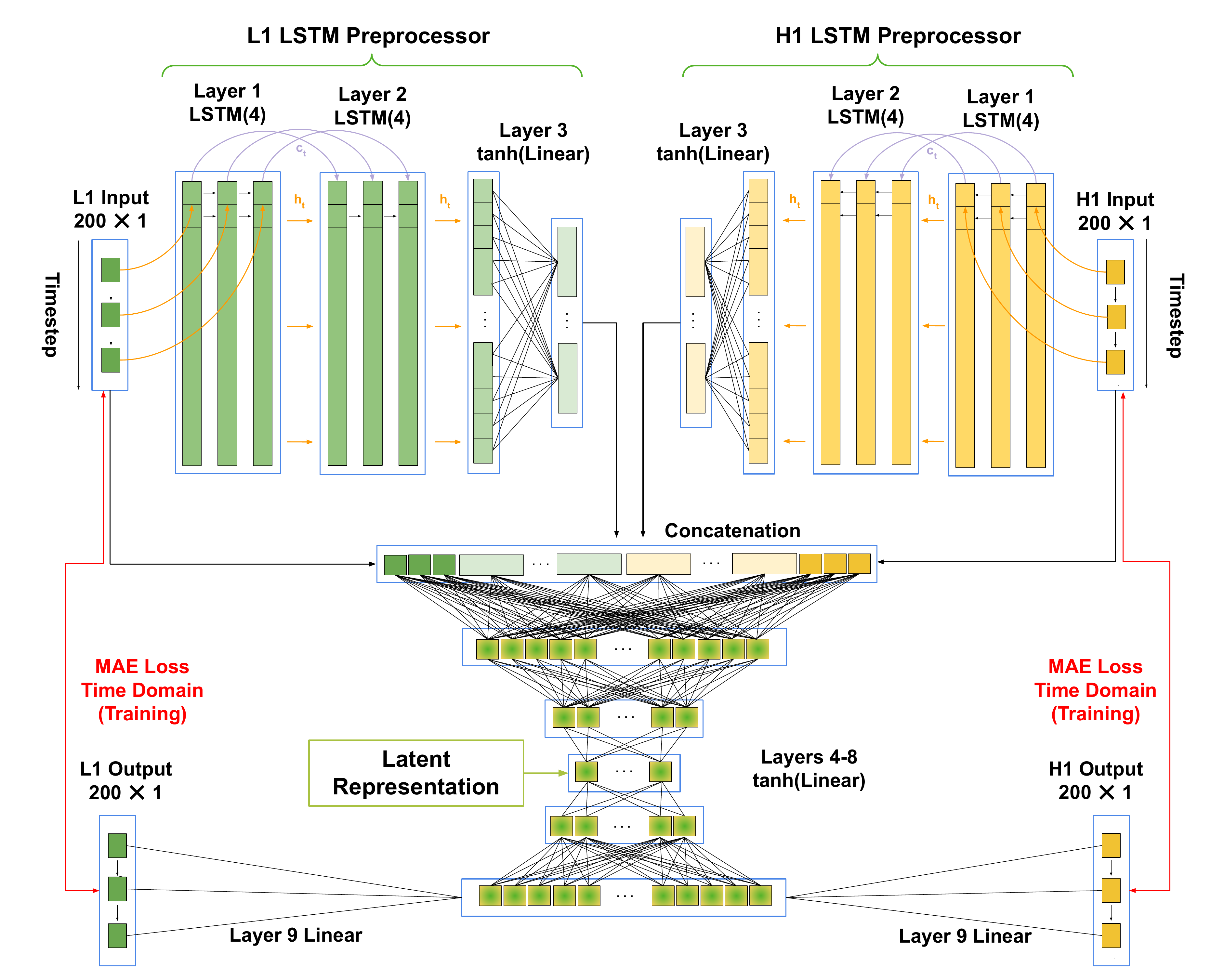}
\caption{Graphical illustration of LSTM autoencoder, preserving the traditional encoder and decoder structure, which allows for a reconstruction loss detection statistic.}
\label{fig:LSTM_AE}
\end{figure}

\begin{figure}[htb]
\centering
    \includegraphics[width=1\textwidth]{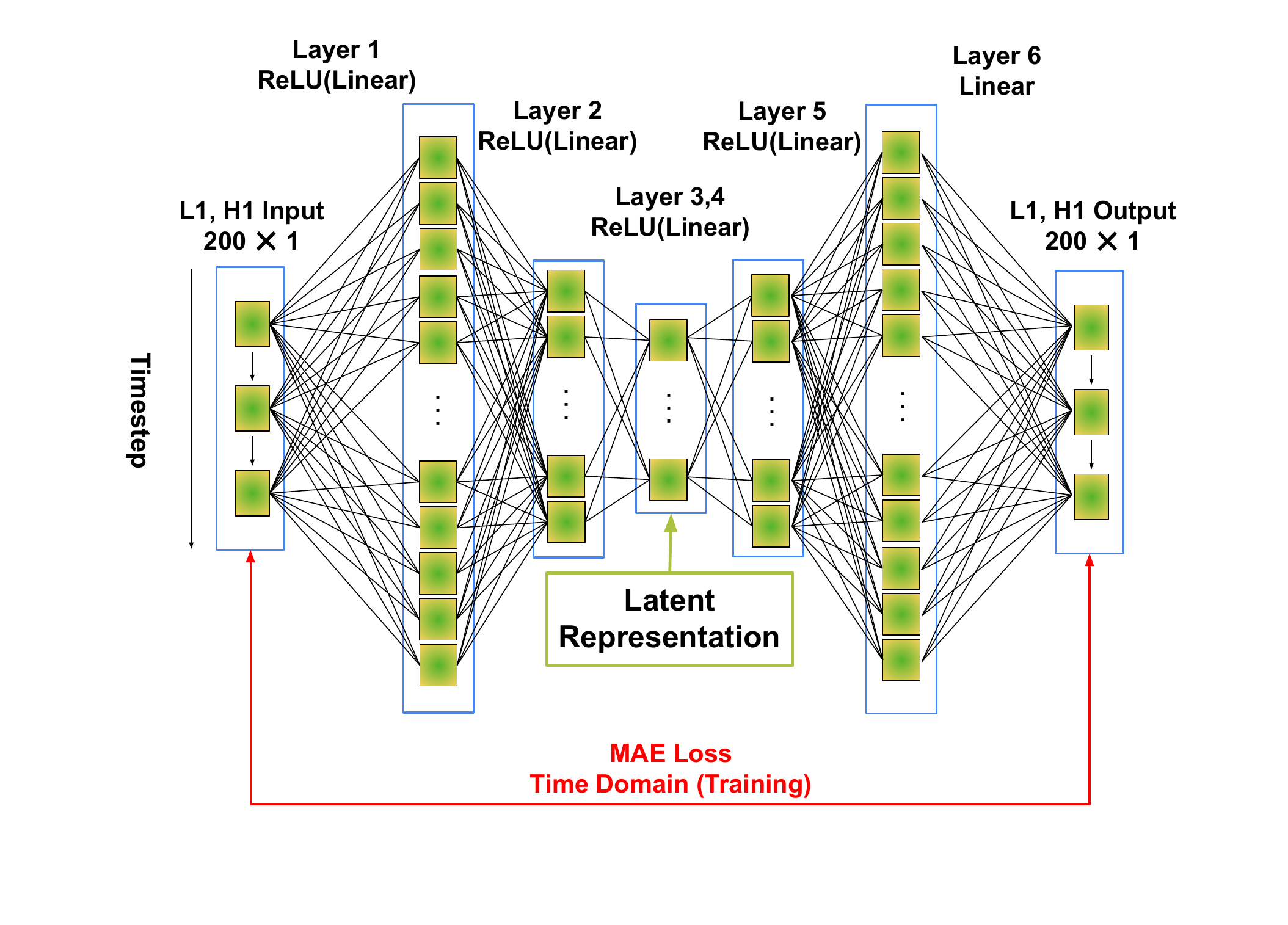}
\caption{Graphical illustration of DNN autoencoder, preserving the traditional encoder and decoder structure, which allows for a reconstruction loss detection statistic.}
\label{fig:DENSE_AE}
\end{figure}

\subsection{Data samples}
\label{sec:data}
\begin{figure}[tb]
\centering
\includegraphics[width=0.7\textwidth]{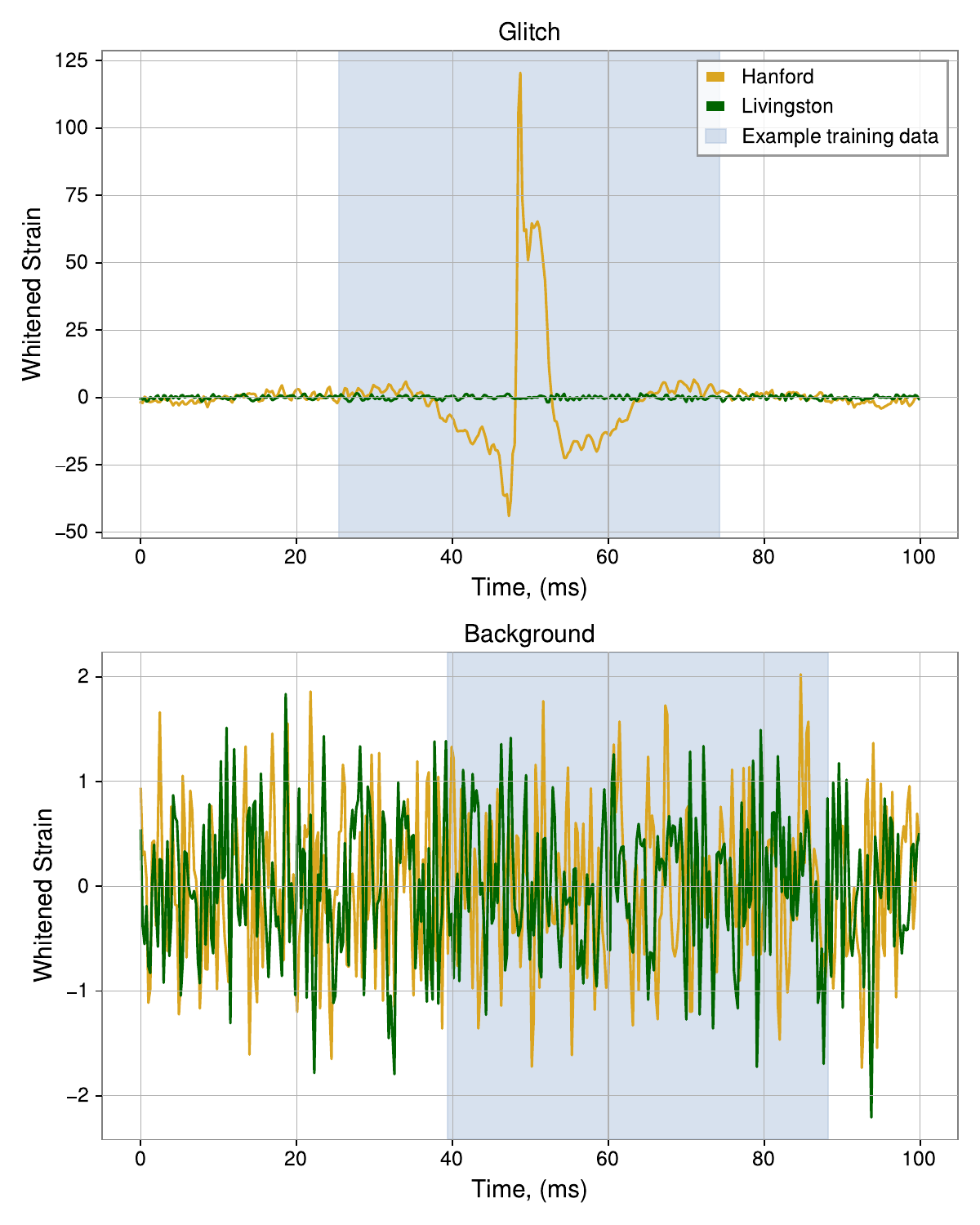}
\caption{Example of GWAK classes: glitch (top) and background (bottom) strains. The light blue
shading highlights an example region that is passed as input to the autoencoders for training.}
\label{fig:background_classes}
\end{figure}

\begin{figure}[tb]
\centering
\includegraphics[width=0.49\textwidth]{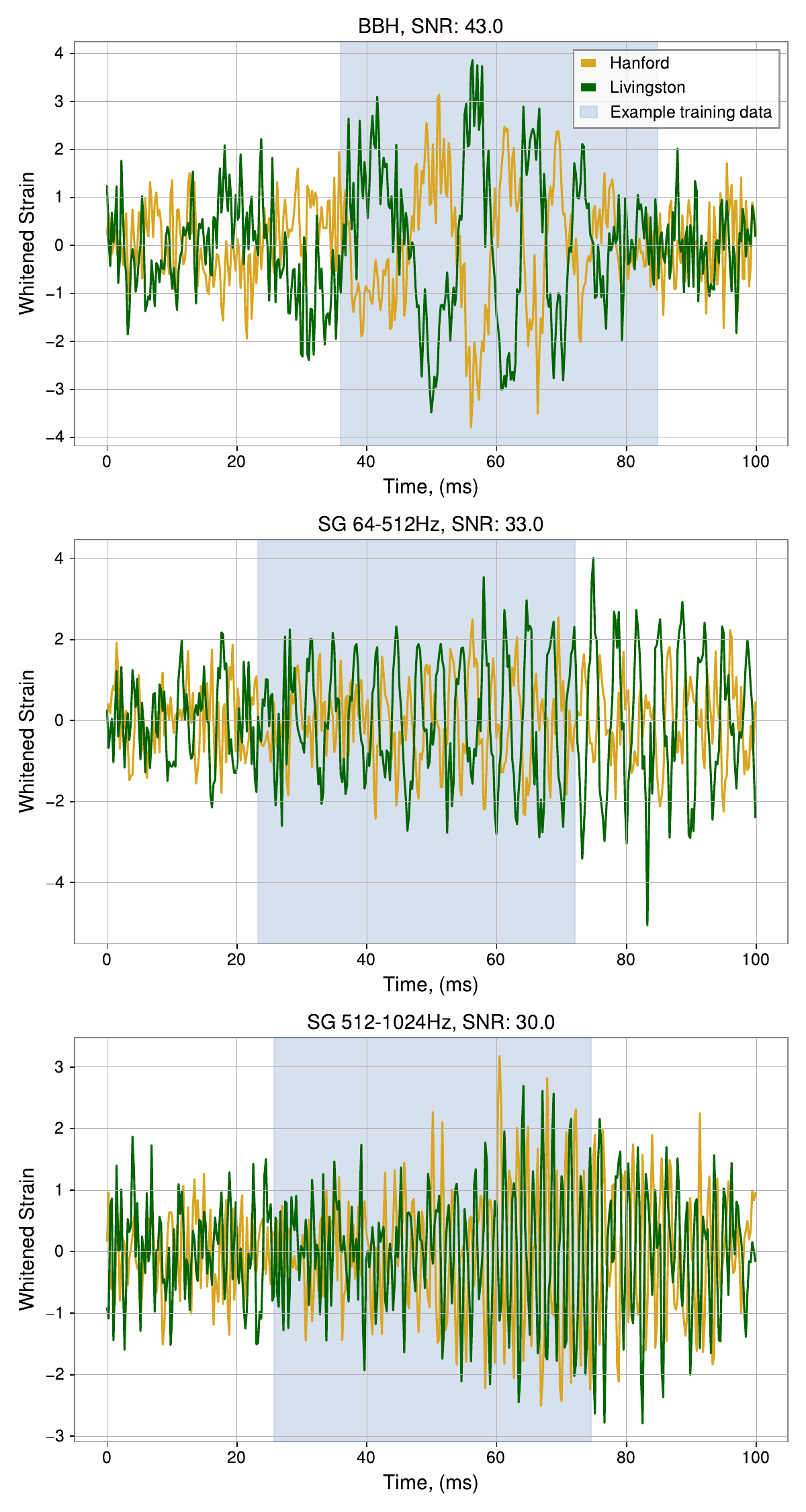}
\includegraphics[width=0.49\textwidth]{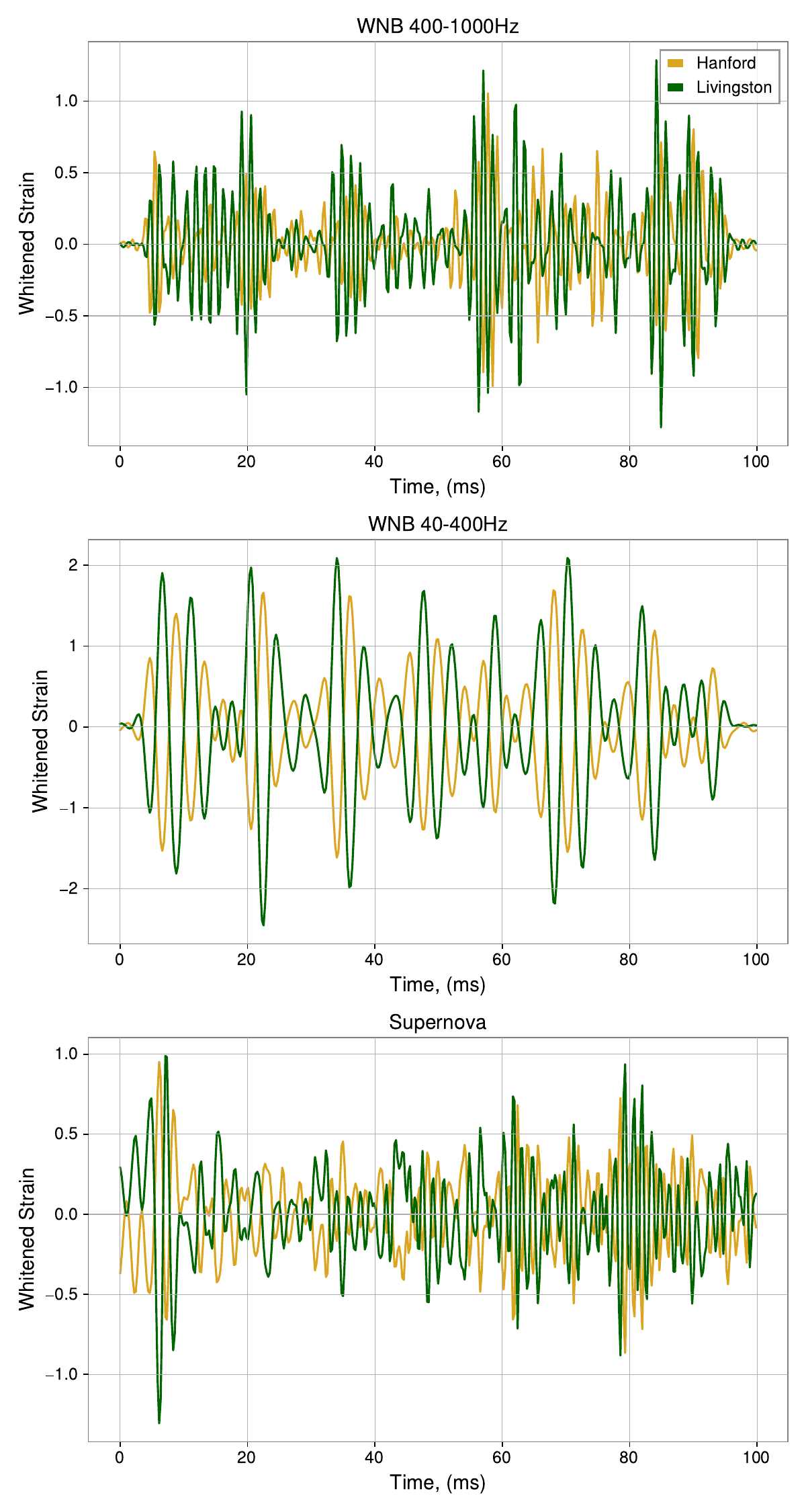}
\caption{Example of signal-like classes: BBH (top left), WNB (top, middle right), sine-gaussian (middle, bottom left) and Supernova (bottom right) strains. The light blue shading highlights an example region that is passed as input to the autoencoders for training.}
\label{fig:signal_classes}
\end{figure}

The dataset used in this study was collected by the LIGO Hanford~(H1) and LIGO Livingston~(L1)~\cite{TheLIGOScientific:2014jea} gravitational-wave detectors during the first half of the third observing run (O3a), which took place between 1st April 2019 and 1st Oct 2019. We specifically used publicly available data between GPS times of 1238166018 and 1238170289, right at the beginning of the run. 
Next, the time-series data were downsampled from 16384\,Hz to 4096\,Hz, and processed to remove and create a separate dataset of transient instrumental artifacts (glitches) using the excess power identification algorithm Omicron~\cite{Robinet:2020lbf}.  We used Q$_{min} = 3.3166$, Q$_{max} = 108$, and f$_{min} = 32$ for the Omicron algorithm.
We then took 4\,s segments of the data without noise artifacts to serve as the baseline for the injection of signals.
As such, we created five different classes of data as proxies for signals and background signatures:
\begin{itemize}
    \item \textbf{Binaries} -- Simulated BBH signals using IMRPhenomPv2~\cite{PhysRevD.82.064016, PhysRevD.93.044006, PhysRevD.93.044007} injected into the real background noise, as shown in Fig.~\ref{fig:signal_classes}(top left). The simulation parameters are given in Table~\ref{table:priors}.
    \item \textbf{Background} -- Background from O3a with DQsecDB ~\footnote{\url{https://git.ligo.org/computing/dqsegdb/client}} state flag \verb!DCS-ANALYSIS_READY_C01:1! applied and excess power glitches~\cite{Robinet:2020lbf} and known GW-events removed, as shown in Fig.~\ref{fig:background_classes}(bottom).
    \item \textbf{Sine-Gaussian 64 -- 512 Hz} -- Generic low frequency signal model used to simulate generic GW sources, as shown in Fig.~\ref{fig:signal_classes}~(middle left).
    \item \textbf{Sine-Gaussian 512 -- 1024 Hz} -- Generic high frequency signal model used to simulate generic GW sources, as shown in Fig.~\ref{fig:signal_classes}~(bottom left).  
    \item \textbf{Glitches} -- Transient instrumental glitches (often of unknown origin) flagged by Omicron~\cite{Robinet:2020lbf} as having excess power, as shown in Fig.~\ref{fig:background_classes}~(top).
\end{itemize}
To create samples of BBH and SG signals, we used numerical simulations to generate $h_+$ and $h_\times$ polarization modes.
We then sampled sky localizations uniformly in the sky, projected the polarization modes onto the sky location, and injected the projected modes into the two LIGO detectors.
\noindent{}

The BBH sample is generated with the parameters and priors~\cite{alex_nitz_2020_3993665} as shown in Table~\ref{table:priors}, taken from \texttt{bilby} processing spins BBH prior, and the sine-Gaussian samples are generated with the parameters and priors as shown in Table~\ref{table:priors}.

\begin{table}[tb]
\centering
\begin{tabular}{ l|l|ccc }
\hline
& Parameter & Prior & Limits & Units\\
\hline
\multirow{9}{*}{\rotatebox[origin=c]{90}{BBH}}  & $m_1$ & - & $(5, 100)$ & $M_{\odot}$ \\
&$m_2$ & - & $(5, 100)$ & $M_{\odot}$ \\
&Mass ratio $q$ & Uniform & $(0.125, 1)$ & - \\
&Chirp mass $M_c$ & Uniform & $(25, 100)$ & $M_{\odot}$\\
&Tilts $\theta_{1,2}$ & Sine & $(0, \pi)$ & rad.\\
&Phase $\phi$ & Uniform & $(0, 2\pi)$ & rad.\\
&Right Ascension & Uniform & $(0, 2\pi)$ & rad.\\
&Declination $\delta$ & Cosine & $(-\pi/2, \pi/2)$ & rad.\\
\hline
\multirow{8}{*}{\rotatebox[origin=c]{90}{sine-Gaussian}} 
&$Q$ & Uniform & $(25, 75)$ & - \\
&Frequency & Uniform & $(64, 512)$ and $(512, 1024)$ & Hz \\
&Phase $\phi$ & Uniform & $(0, 2\pi)$ & rad.\\
&Right Ascension & Uniform & $(0, 2\pi)$ & rad.\\
&Declination $\delta$ & Cosine & $(-\pi/2, \pi/2)$ & rad.\\
&Eccentricity & Uniform & $(0, 0.01)$ & -\\
&$\Psi$ & Uniform & $(0, 2\pi)$ & rad.\\
\hline
\end{tabular}
\caption{Sampling parameters and priors for BBH (top) and sine-Gaussian (bottom) injections.}
\label{table:priors}
\end{table}

We employed a series of digital signal processing techniques to prepare data for training autoencoders in the context of GW detection. Specifically, we first applied a whitening filter to normalize the data with respect to one hour of surrounding background data. This filter effectively suppressed frequency regions dominated by noise and reduced effects from spectral lines~\footnote{\url{https://gwosc.org/s6speclines/}}~\footnote{\url{https://dcc.ligo.org/LIGO-T1500415/public}}. 
Moreover, we implemented a band-pass filter within the frequency range of 30--1500\,Hz to further attenuate noise outside of the most sensitive frequency range of GW instruments. After applying these filters, we removed 1\,s intervals from each end of the data samples to eliminate any edge effects from preprocessing. The remaining 2\,s samples, each containing either an injected signal, pure background, a low/high frequency sine-Gaussian, or a glitch artifact, were used to generate training data.

To obtain a set of windows suitable for training, we extracted 200 data-points (total duration of 50\,ms sampled at 4096\,Hz) from each sample. Our experimentation revealed that training the autoencoders with input lengths greater than 200 data-points would result in degraded performance, due to the difficulty to capture long-term behavior in an LSTM, especially on an evolving signal. While reducing the number of data-points below 200 could enhance computational efficiency, it reduces the autoencoder's ability to learn the evolution of a shorter-duration signal.

To optimize the data processing and facilitate learning by the network, the data are normalized to have a standard deviation of one on a sample-per-sample basis. This normalization was undertaken mainly for the reason that the neural networks struggled to learn with unnormalized samples. The strongest example of this is with the glitch dataset, where strain magnitude can reach amplitudes hundreds to thousands of times above the background. 

For each of the non-coherent classes (background and glitch), the dataset is split into three parts: 80,000 training samples (80\%), 10,000 validation samples (10\%), and 10,000 test samples (10\%). 
For each signal class (BBH, low frequency SG and high frequency SG), we generate 5 sub-datasets, each with a specified SNR injection range, as shown in Fig.~\ref{fig:signal_loss}. Each of these sub-datasets has an identical splitting procedure: 80,000 training samples (80\%), 10,000 validation samples (10\%), and 10,000 test samples (10\%).
The training and validation datasets were used for the autoencoders to create loss values on which to update and score the networks respectively. The test samples were used for the recreation plots, as in Fig.~\ref{fig:recreation}, as well as to show the GWAK feature space, Fig.~\ref{fig:trained_GWAK}.

Signal events needed to build the GWAK space were created injecting simulated GWs on top of artifact-free detector noise.
This provides an analogous situation to a real GW, in which the strain from the incoming wave is recorded in combination with the detector noise.
We do not explore the case of coincident detector aritfacts with transient astrophysical signals.
The injection of signal also accounts for the difference in GW time-of-arrival at each detector owing to light travel time from the sky localization of the signal, which is significant at a sampling rate of 4096\,Hz, corresponding to a maximum of 40 samples.

\subsection{Autoencoder Training}

To train an autoencoder, the input sequence is passed through the encoder and the decoder, and the model is optimized to minimize the reconstruction error between the input and the output sequence. 
Once the model is trained, it can be used to identify data points that deviate from the normal pattern by comparing the reconstruction error of new data with a threshold.

\begin{figure}[tb]
\centering
\includegraphics[width=0.9\textwidth]{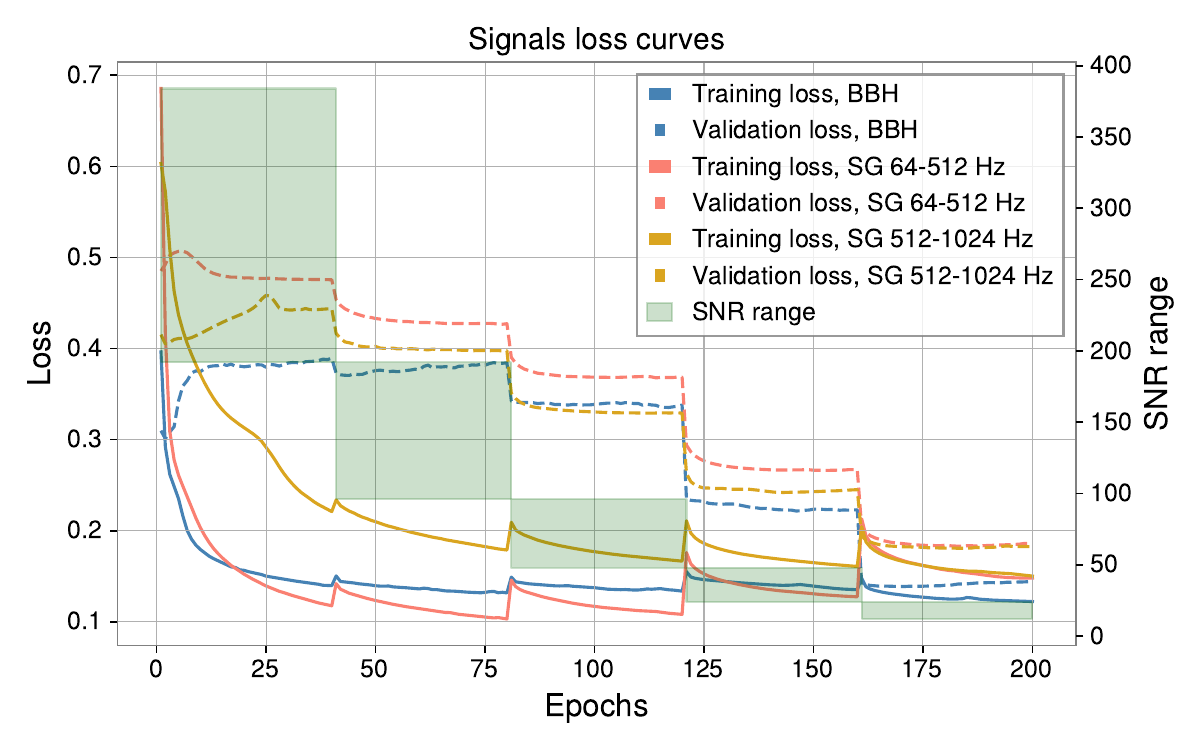}
\caption{
Autoencoder training and validation losses for signal classes, using curriculum learning to progressively reduce validation loss. 
The validation loss for each training SNR range is computed on the validation data from the last SNR step. 
In solid/dashed colours are the training/validation losses for BBH~(blue), SG 64--512\,Hz~(salmon) and SG 512--1024\,Hz~(dark yellow). 
A light green shaded region depicts the SNR range for each step of training, spanning the range of injected SNR corresponding to each curriculum. 
For example, the first curriculum contains injections in the SNR range $[192, 384]$. 
At the transition between curricula, we see a small spike in the training loss, especially for the SG 512--1024\,Hz model. 
This is due to the transition to a ``more difficult'' training dataset, as the lower SNR leads to a less distinguishable signal. 
With the transition between curricula, we see a sharp drop in validation loss over the course of a few epochs. This is due to the fact that the validation set is maintained for each autoencoder through training as belonging to the lowest - $U[12, 24]$ SNR range. 
With the transition to a new curriculum with a lower SNR range, the training data for that curriculum will more closely match the validation set, explaining the rapid drop.
}
\label{fig:signal_loss}
\end{figure}

\begin{figure}[tb]
\centering
\includegraphics[width=1.0\textwidth]{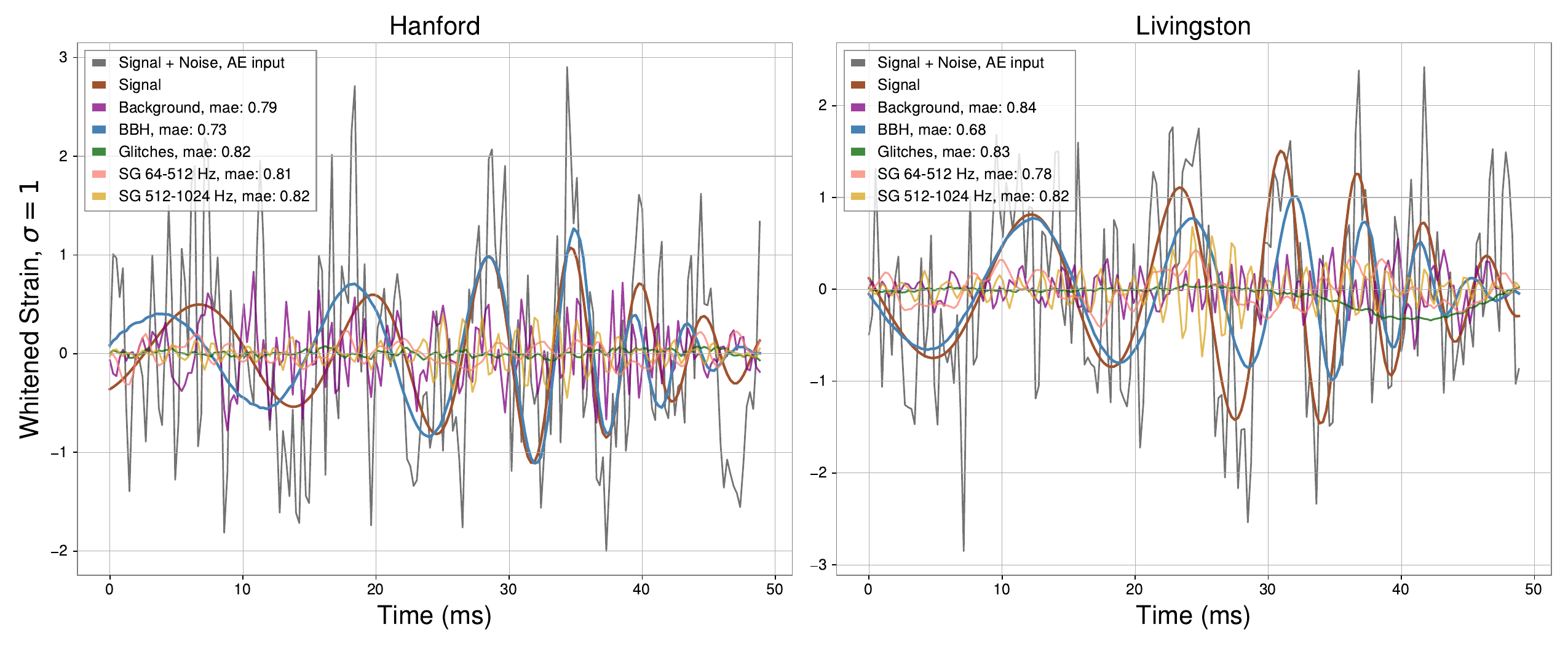}
\caption{Example of recreation on injected BBH signal, with the noise-less template also shown. The recreation of the BBH autoencoder~(in blue) follows closely the original signal injection~(in brown) as expected, since it was trained to reconstruct the noiseless input. While background~(in purple), glitches~(in green), SG 64--512\,Hz and  SG 512--1024\,Hz clearly fail to reconstruct the injected BBH signal, as expected.}
\label{fig:recreation}
\end{figure}

To train our five autoencoders, each corresponding to one of the data classes, we used two different schemes. To train the glitch and background classes, we use the same dataset for all 200 epochs, the Adam~\cite{kingma2017adam} optimizer, and Mean Absolute Error (MAE) loss, computed between the autoencoder input and autoencoder reconstruction. To train the binary black hole, sine-Gaussian low frequency and high frequency classes, we used a curriculum scheme. Over 200 epochs, we used 5 different datasets, each with identical injections but different uniform SNR priors: $U[192, 384]$ (epochs 1-40), $ U[96, 192]$ (epochs 41-80), $U[8, 96]$ (epochs 81-120), $U[24, 48]$ (epochs 121-160), $U[12, 24]$ (epochs 161-200), as shown in Fig.~\ref{fig:signal_loss}. Here, we used the Adam optimizer, reset after each step of the curriculum, and computed error between the autoencoder output of a noisy input (injection into the real background with specified SNR) and the ``clean,'' noise-less template. This was intended to train the autoencoders to learn to reconstruct the signal itself, without any noise. To compute the validation loss at each epoch, we used a subset of the $U[12, 24]$ SNR dataset for each curriculum. This was intended to provide a fair computation of the validation loss across each curriculum.
The MAE loss $\mathbf{L}$ between original data $\mathbf{D}$ and reconstruction $\mathbf{R}$ is given by
$$\mathbf{L} = \frac{1}{N} \frac{1}{2} \frac{1}{200} \sum_{i=1}^N \sum_{j=1, 2} \sum_{k=1}^{200} |\mathbf{D}_i [j, k] - \mathbf{R}_i [j, k] |$$
$\mathbf{D}_i [j, k]$ represents the i-th sample of the original data, taken from the j-th detector at the k-th timestep, and likewise $\mathbf{R}_i [j, k]$ represents the i-th sample of the autoencoder output, taken from the j-th detector at the k-th timestep. The loss curves are shown in Fig.~\ref{fig:signal_loss}.
The example of autoencoder reconstruction obtained with pre-trained autoencoders is shown in Fig.~\ref{fig:recreation}, and recreation samples of other training classes are shown at the end of the paper.

\subsection{Feature extraction}
\label{sec:featextract}
While the MAE loss was used in training, at evaluation time we opted for a frequency-domain based features, computed between the input and autoencoder output. The intuition for choosing to compute features in the frequency domain is based on the fact that signal autoencoders were trained with clean, noiseless targets. Since the reconstruction is a noise-less signal, and one 50\,ms window of a signal does not exist in a wide range of frequencies, the reconstruction will be a localized peak in frequency space. On the contrary, the original signal, especially one of low SNR, will contain both the narrow-bandwidth frequency feature along with noise distributed approximately evenly throughout the entire frequency range. When computing the MAE between the original input and reconstructed output, the presense of high frequency noise will inflate the loss, since the autoencoder, by design, does not fit noise. To bypass this, we chose to compute a dot product in frequency space. In the frequency regime where the true signal exists, both original and reconstructed signals will be similar, and as such the dot product will yield a high value. In the other frequency regimes where the signal is not present, the reconstructed signal is close to zero, and as such these noisy contributions get removed.

We choose two features per autoencoder to be the following: Let $H_O, L_O, H_R, L_R$ correspond to the original Hanford and Livingston signals and reconstructed Hanford and Livingston signals respectively, from a single autoencoder. Each are 200-datapoint segments, sampled at 4096 Hz.\,We then take the Fourier transform of each, yielding $\widetilde{H_O}, \widetilde{L_O}, \widetilde{H_R}, \widetilde{L_R}$. The two features, per autoencoder, are $|\widetilde{H_O} \cdot \widetilde{H_R} |$ and $|\widetilde{L_O} \cdot \widetilde{L_R} |$. We also used a general ``frequency space correlation'' feature to compliment the Pearson correlation ~\ref{sec:pearson}, defined by $|\widetilde{H_O} \cdot \widetilde{L_O} |$. $|\widetilde{A} \cdot \widetilde{B}|$ represents the magnitude of the dot product of two complex vectors, namely the Fourier transforms of $A$ and $B$. Training an autoencoder to optimize directly on the $|\widetilde{H_O} \cdot \widetilde{H_R} |$, $|\widetilde{L_O} \cdot \widetilde{L_R} |$ features proved to be too simple of a task. Each autoencoder would consistently learn the largest feature in Fourier space, leading autoencoders generalizing too well, i.e. not being specific enough to their respective training class. By scoring the network's performance in the time domain, i.e. with MAE between the input and reconstructed output, the network must accurately recreate more details, such as temporal offsets, signal evolutions, as well as the corresponding frequency components, forcing the specificity to the class. Finally, by training with MAE, it provides us with a nice visual picture of the autoencoder output which we can easily compare against the input, but this would not necessarily be the case of using the frequency-domain features directly for loss. 

\subsection{Pearson cross-correlation}
\label{sec:pearson}
To derive a comprehensive metric for inference, we incorporated information regarding the cross-correlation between the two detector sites, in conjunction with the GWAK information. Given that any astrophysical signal will invariably manifest in both detector sites, the correlation of the measured strains is of critical significance for the signal search procedure. Although we utilized information from both detectors during the GWAK space training phase, we opted to directly incorporate cross-correlation information in our final metric.

To accomplish this, we employed the Pearson correlation coefficient~\cite{pearson}. Specifically, we computed the Pearson correlation coefficient between the Hanford and inverted Livingston sites by selecting the maximum correlation coefficient from all possible time shifts for a 200 datapoint window. Since the physical separation between the detectors is about 3000\,km, corresponding to a time of flight of 10\,ms, we must iterate over all possible temporal shifts within 10\,ms to contain the correct time delay. At a sampling rate of 4096\,Hz, this corresponds to a shift of 40 datapoints in either direction. This iteration over all possible time shifts is an advantage of the Pearson correlation coefficient over the frequency-domain correlation coefficient, and as such we chose to include both in our GWAK space. The source of this time delay is due to the sky location of the gravitational wave source.  The Livingston detector is inverted to account for the detectors' relative orientations~\cite{Abbott_2016}. The Pearson correlation coefficient is a widely used statistical measure that provides a measure of the strength of a linear relationship between two variables, in this case, the strain measurements from the two sites. The coefficient ranges from $-$1 to 1, with values close to 1 indicating a strong positive correlation, values close to $-$1 indicating a strong negative correlation, and values close to 0 indicating a lack of correlation between the two variables.

Given two detector streams, the Pearson cross-correlation at time t was computed via
$$\mathbf{P} = \max_{\Delta} \frac{\sum_{k=t-100}^{t+100} (H_k - <H>) \cdot ( L_{k- \Delta} - <L>) \cdot (-1)}  {\sqrt{ \left( \sum_{i=t-100}^{t+100} (H_i-<H>)^2 \right)\left( \sum_{j=t-100-\Delta}^{t+100-\Delta} (L_j-<L>)^2 \right)} },$$
$$<H> = \frac{1}{200} \sum_{i=t-100}^{t+100} H_i, \:\: <L> = \frac{1}{200} \sum_{j=t-100-\Delta}^{t+100-\Delta} L_j$$

The presence of a multiplicative factor of $-$1 serves as the inversion of the Livingston detector due to the relative orientation. The range of $\Delta$ is within the maximum time of flight in units of datapoints, so $\Delta \in [-40, 40]$.
In addition to the autoencoder frequency domain features, this yields $5$ (autoencoders) $\cdot 2$ (features per autoencoder) $ + 1$ (pearson) $ + 1$ (frequency-domain correlation) $ = 12$ overall features.

\subsection{Artificial background with timeslides}
\label{sec:timeslides}
To create background data that we used a technique called timeslides. This involved temporally shifting the data from one detector relative to another by at least 10\,ms, the light travel time between detectors, guaranteeing that there is no astrophysical correlation present. As a background dataset for \ref{sec:lincomb}, we computed 8 hours worth of timeslides. To evaluate our entire algorithm and report false alarm rates for detections, we computed 1 year worth of timeslides.

\subsection{Linear combination optimization}
\label{sec:lincomb}
To construct a final metric which combines the information from the above 12 features, we opted for a linear combination of those values to produce one final metric value. 
Given that we are trying to generalize to unknown anomalous signals, opting for a simple linear model aims to reduce any bias towards known signal regions. 
To find optimal values for the parameters of the linear classifier, we used a simple Linear Support Vector Machine~(SVM), which aims to optimize the binary classification of background and signal classes. 
The classification by a Linear SVM is simply given by $\vec{W}^T \vec{X} + b$, where $\vec{W}$ represents the learned weight vector, of the same dimensionality as the GWAK vector $\vec{X}$, which contains the 12 GWAK features. $b$ represents a bias term.
For the background dataset, we used 8 hours worth of timeslides as described in \ref{sec:timeslides}.
Also using this stretch of timeslides, we computed a set of normalization coefficients. For each of the 12 features, these were the mean value and standard deviation of that feature across 8 hours of analysis. These were used as to rescale each feature to mean zero and standard deviation one independently. This helped with training the linear classifier, as while the Pearson correlation coefficient is O(1), the frequency domain coefficients can be O(1000).
For the signal dataset, we generated a new dataset, comprising of 6 classes - BBH, SG (64 -- 512 Hz), SG ( 512 -- 1024Hz), low-frequency white noise burst (40-400Hz), high frequency white noise burst (400-1000Hz), and supernova ~\cite{Powell_2021} (85 $M_{\odot}$ progenitor mass, SFHo equation of state), each with a SNR prior of $U(10, 100)$.

\begin{figure}[htb]
\centering
\includegraphics[width=1.0\textwidth]{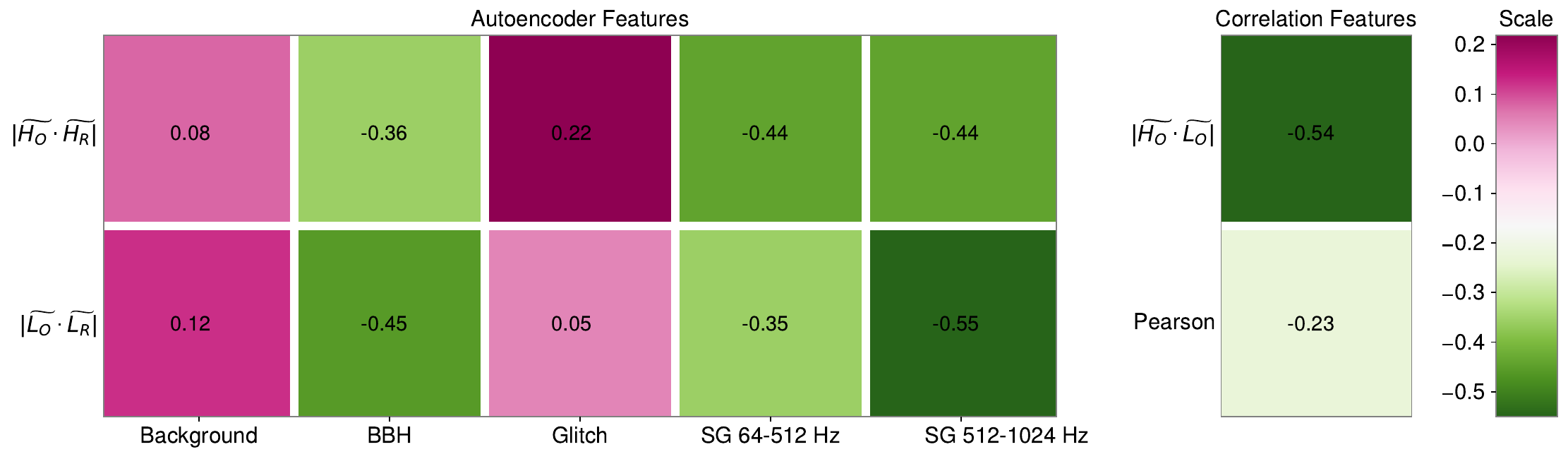}
\caption{Learned coefficients of a linear combination of the 12 GWAK features. As expected, the features corresponding to the signal autoencoders have more negative values, as well as the correlation features, as a higher correlation score is more representative of an astrophysical event. While background and glitch features have more positive values since a stronger relationship to those classes will be less indicative of the signal.}
\label{fig:fm_weights}
\end{figure}

We trained this linear fit for 5000 epochs, with a learning rate of 0.01 and using the Adam optimizer. 
Fig.~\ref{fig:fm_weights} shows the learned coefficients for each of the 12 GWAK features. This serves as a sanity check that the impact of each feature on the final metric score is as we expect. Signals are classified via a more negative value, so the features corresponding to the signal autoencoders should have more negative values, as well as the correlation features, as a higher correlation score is more representative of an astrophysical event. On the contrary, autoencoders corresponding to non-astrophysical data (background or glitch) should have more positive values, since a stronger relationship to those classes will be less indicative of the signal. These intuitions are all demonstrated via Fig.~\ref{fig:fm_weights}, so we pass this sanity check.

\subsection{Smoothing for the final metric}
\label{sec:smoothing}
As the GWAK algorithm assigns a single metric value to each 50\,ms window, it does not naively carry sensitivity to signals longer then the 50\,ms window. A longer signal will have multiple evaluation points, but to compute FAR only the lowest metric value is taken. To increase sensitivity by specifying the GWAK algorithm to signals greater than one window in length, we convolved the timeseries of final metric evaluations with uniform kernels of varying size, with the idea being that the sensitivity to a given anomalous signal is maximized when the kernel length is of order the signal length. We present the detection efficiency using this method in Fig.~\ref{fig:smoothing}.

\section{Results}
\label{sec:results}

This section describes how our proposed methodology can be used to discover anomalous events. Additionally, we evaluate the effectiveness of our semi-supervised approach in detecting several potential sources of GWs, without using information about these signals during the training phase.

\subsection{Background and glitch mitigation}
As the goal of this method is to identify anomalous signals in the background, the mitigation of non-astrophysical data being identified as signal-like improves sensitivity to real signals by reducing the corresponding false alarm rate. In particular, detector artifacts or ``glitches'' can often pose a problem as they can have signal-like morphology \cite{LIGOScientific:2016gtq, Cabero:2019orq} as well as possess high SNR in a single detector. This serves as the motivation for training a glitch autoencoder, as it should be able to recognize single detector artifacts reliably. Upon ``recognition'' of a glitch, via the glitch autoencoder frequency domain features, an automatic down-weight will be given to the event in question. In addition, as glitches are uncorrelated between detectors, occurring locally, the use of both correlation features also helps to mitigate false alarms caused by glitches. As there is no correlation between observations in one detector stream and another within the maximum light travel distance during a glitch, those features will correspondingly have small values, and similarly down-weight the glitch to have a less signal-like score.

\subsection{Anomaly metric}

\begin{figure}[htb]
\centering
\includegraphics[width=0.99\textwidth]{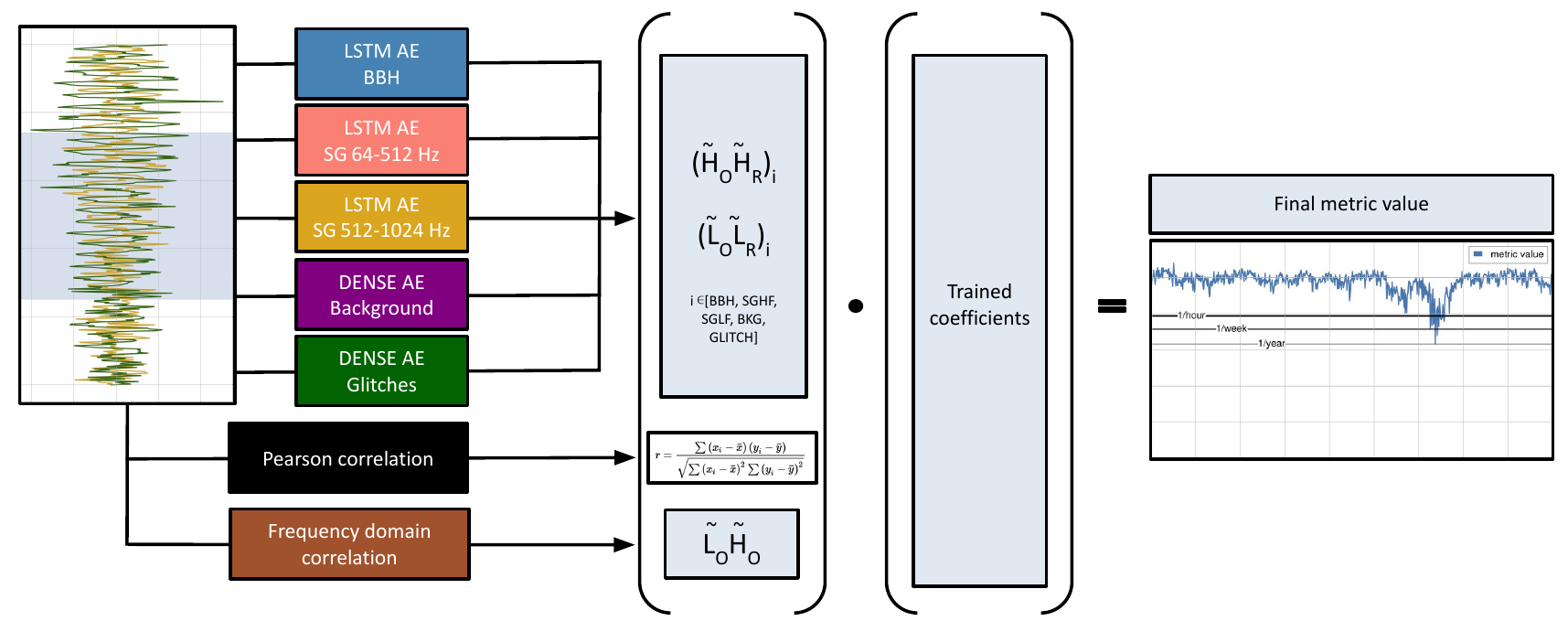}
\caption{Example of metric calculation on a hypothetical event. The event is reconstructed with each of the 5 pre-trained autoencoders. Pearson and frequency domain correlation are computed on the given input and then each of the values is multiplied with a corresponding coefficient which arises from the pre-trained linear metric. The sum of all the features multiplied by their coefficients is referred to as the final metric and is used to make a decision on if the event is signal-like.}
\label{fig:metric_scheme}
\end{figure}

We employ the linear combination method outlined in Sec.~\ref{sec:lincomb}, which includes the two frequency-domain features per autoencoder, the frequency-domain correlation, and the Pearson cross-correlation presented in Sec.~\ref{sec:pearson}.    
The example of a full GWAK pipeline is shown in Fig.~\ref{fig:metric_scheme}.

\begin{figure}[b]
\centering
\includegraphics[width=0.99\textwidth]{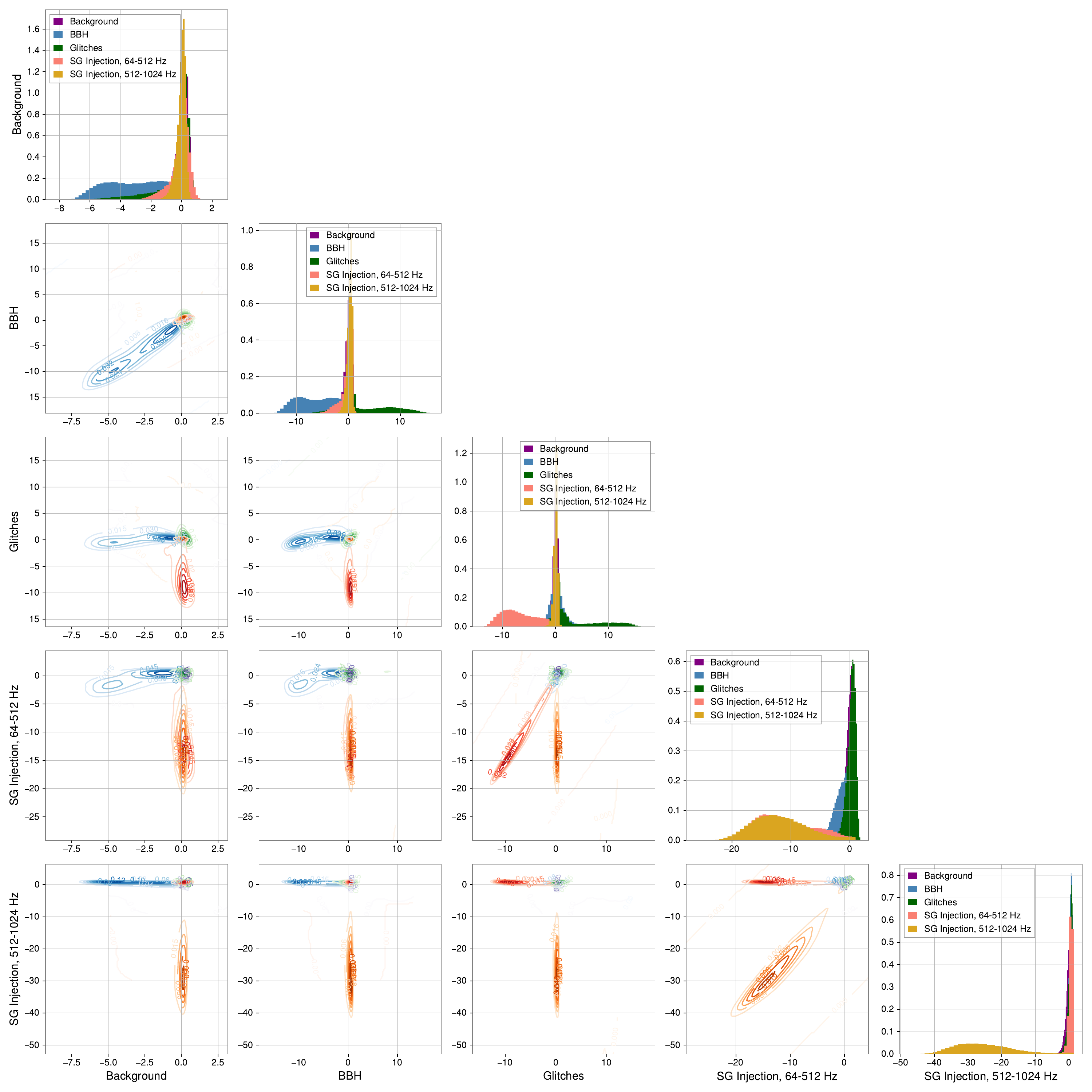}
\caption{Trained GWAK space on 5 signal classes: BBH, background, glitch, and low and high frequency Sine-gaussian; demonstrating that regions of the GWAK space correspond to different event classes}
\label{fig:trained_GWAK}
\end{figure}

The resulting coefficients of the linear classifier, physically describing a hyperplane, are then used to project points from the 12-dimensional feature space down to a one-dimensional space via the dot product, or geometrically the perpendicular distance from that point to the classifying hyperplane. This one-dimensional real number is our final metric value. 
Since our classifier was trained to predict 0 for signals and 1 for backgrounds, a more negative final metric value corresponds to a more ``signal-like'' or louder input, whereas a less negative/positive metric value indicates background or no signal of interest.
Once we compute the final metric value for a hypothetical signal, we would like to know the corresponding false alarm rate. This corresponds to the frequency that a non-astrophysical input (glitch or otherwise) from the gravitational-wave detectors would lead to a trigger of the same significance as the hypothetical signal.
The 12 features multiplied by the corresponding coefficient and grouped by a corresponding autoencoder are shown in Fig.~\ref{fig:trained_GWAK}. We show each of the five data classes in this space to highlight how strongly the signals are separated and grouped in this new, learned GWAK space.

\subsection{Evaluation on Core-Collapse Supernova}
We use existing core-collapse supernovae simulations to see how our approach extends to anomalous signals which GWAK has not seen during training. We use ~\cite{Powell_2021} (85 $M_{\odot}$ progenitor mass, SFHo equation of state).
On the top left of Fig.~\ref{fig:3_panel_plots}, the evaluation of GWAK axes and Pearson correlation with time and on the top right total metric value and FAR are shown as an example of the algorithm's ``reaction'' to unseen signals. Both BBH and sine-Gaussian losses drop at the time of the signal,  which indicates that the strain at that moment is more signal-like. The Pearson correlation increases indicating a strong correlation between the two detector sides. The FAR at the moment of the event drops to a level consistent with one or fewer events per month, which means that even with strong trigger restrictions, detection of that type of event would be possible with our new proposed algorithm. 

The bottom plot in Fig.~\ref {fig:3_panel_plots} (left) reflects the scan of minimum metric value for different signal-to-noise ratio ranges of the injected core-collapse supernova models. As expected, with the higher signal-to-noise ratio, the total metric is lower resulting in a lower FAR. If the trigger bandwidth would allow for up to 1 false event per hour being triggered, GWAK would be able to detect core-collapse supernova events with $\sim$22 SNR.

\subsection{Evaluation on White Noise Bursts}

\begin{figure}[htb]
\centering
\includegraphics[width=0.49\textwidth]{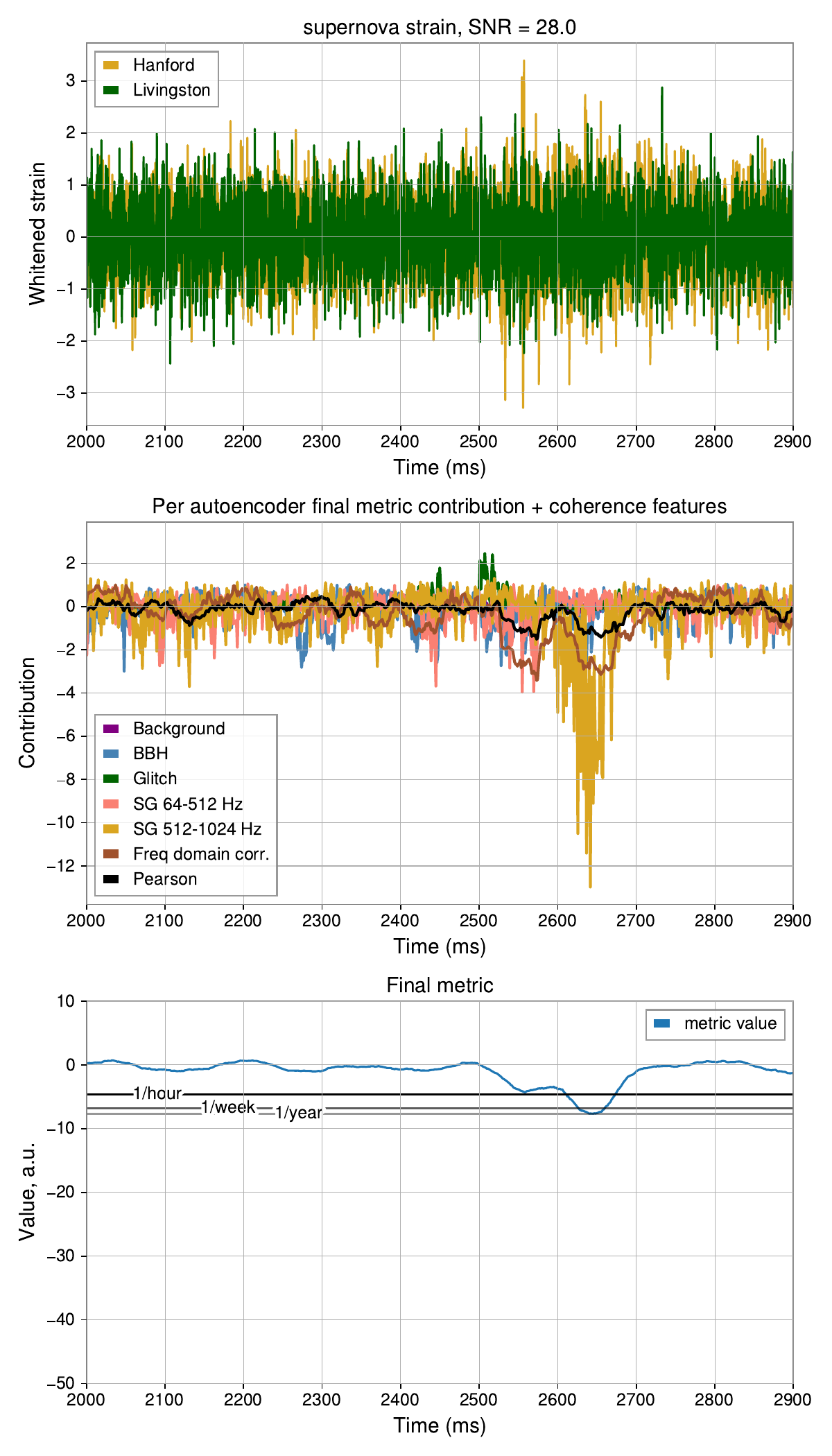}
\includegraphics[width=0.49\textwidth]{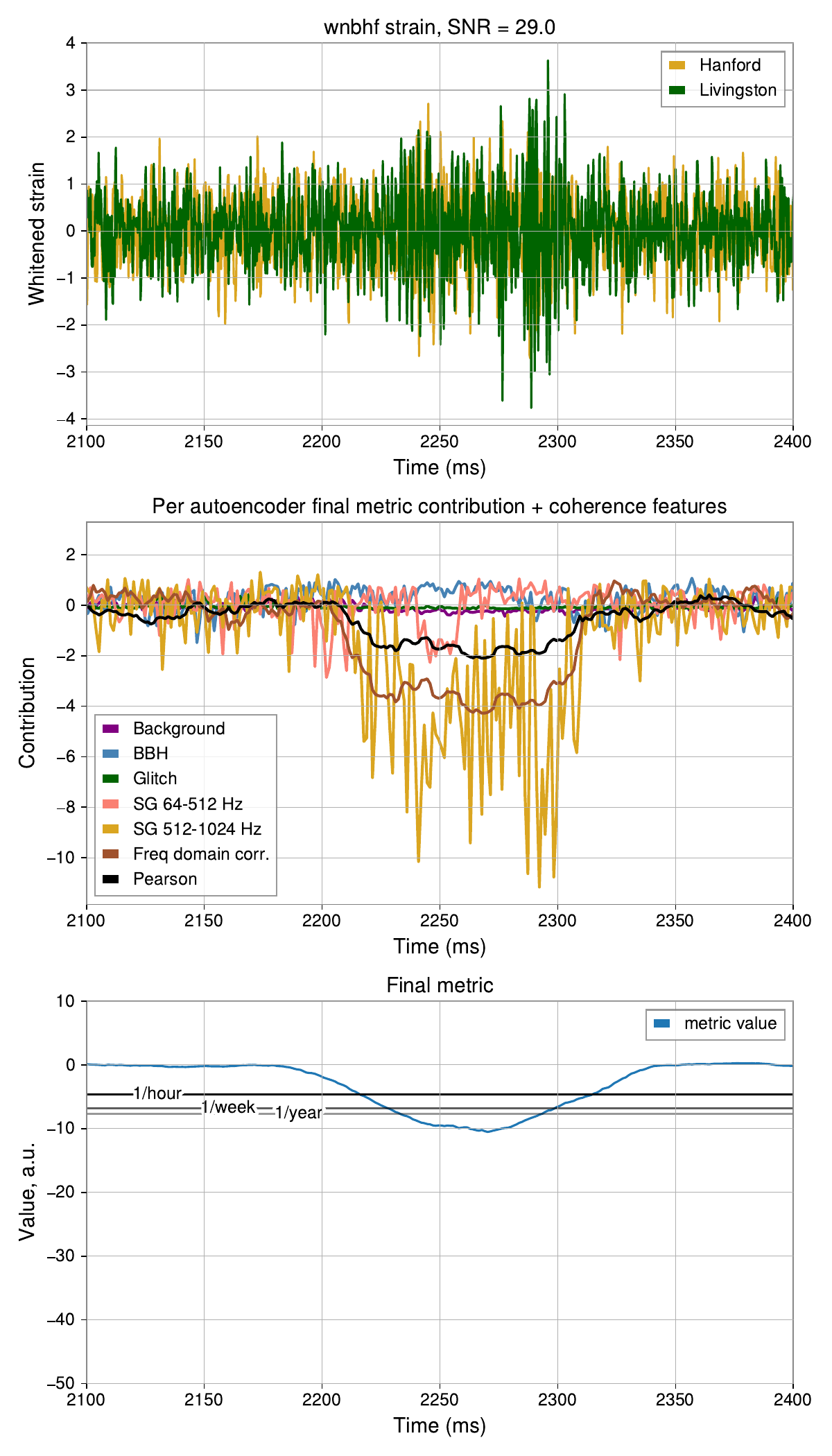}
\caption{Strain (top), GWAK metric response (middle) and final metric response (bottom) for WNB (right) and Supernova (left).
The evaluation of GWAK axes and Pearson correlation with time and on the top right total metric value and FAR are shown as an example of the algorithm's ``reaction'' to unseen signals. Both BBH and sine-Gaussian losses drop at the time of the signal,  which indicates that the strain at that moment is more signal-like. The Pearson correlation increases indicating a strong correlation between the two detector sides. At the moment of the event, the FAR drops to a level consistent with or below one event per year, which means that even with strong trigger restrictions, detection of that type of event would be possible with our new proposed algorithm. }
\label{fig:3_panel_plots}
\end{figure}

Furthermore, we assessed the performance of our method on white-noise bursts, which are signals characterized by the presence of $h^+$ and $h^{\times}$ polarizations that are independent time series of Gaussian noise, which is whitened over a specific frequency range and multiplied by a sigmoid envelope. 
The bandwidth of each injected signal is selected uniformly and randomly from a range spanning 40--400\,Hz and 400--1000\,Hz. 
The duration of the signal is chosen to be 0.1\,s.
Theoretically, these would be the most difficult signals to detect with our algorithm, as their lack of distinctive morphology would render the signal autoencoder features useless. However, as shown in Fig.~\ref {fig:3_panel_plots}, the SG autoencoders were able to generalize to the WNBs.

To evaluate the performance of our algorithm, we generated these signals with SNRs uniformly distributed between 10 and 100. 
The average final metric value and corresponding standard deviation for various SNR ranges are shown in Fig.~\ref{fig:3_panel_plots} (right) and the lines corresponding to different false alarm rates. 

The demonstration of the GWAK algorithm from strain to final metric is shown in Fig.~\ref{fig:3_panel_plots}. Starting with the whitened strain, we split up our data into 200 datapoint windows with a step of 5 datapoints between windows. For each window, the 10 autoencoder features are computed, along with the frequency domain correlation, and Pearson correlation, as described in Sec.~\ref{sec:pearson}. Using the learned SVM weights as shown in Fig.~\ref{fig:fm_weights}, we reduce the 12-dimensional GWAK space to 7 dimensions, and show those values at each timestep in the second panel. For the correlation features, this is simply done by multiplying the value by the corresponding weight. For the autoencoder features, the same is done, but the $|\widetilde{H_O} \cdot \widetilde{H_R} |$ and $|\widetilde{L_O} \cdot \widetilde{L_R} |$ values are combined into a single value via addition. 
Finally, all of those weighted features are summed yielding the final metric value, which is shown in the third panel, along with various false alarm rate thresholds.

\begin{figure}[tb]
\centering
\includegraphics[width=0.8\textwidth]{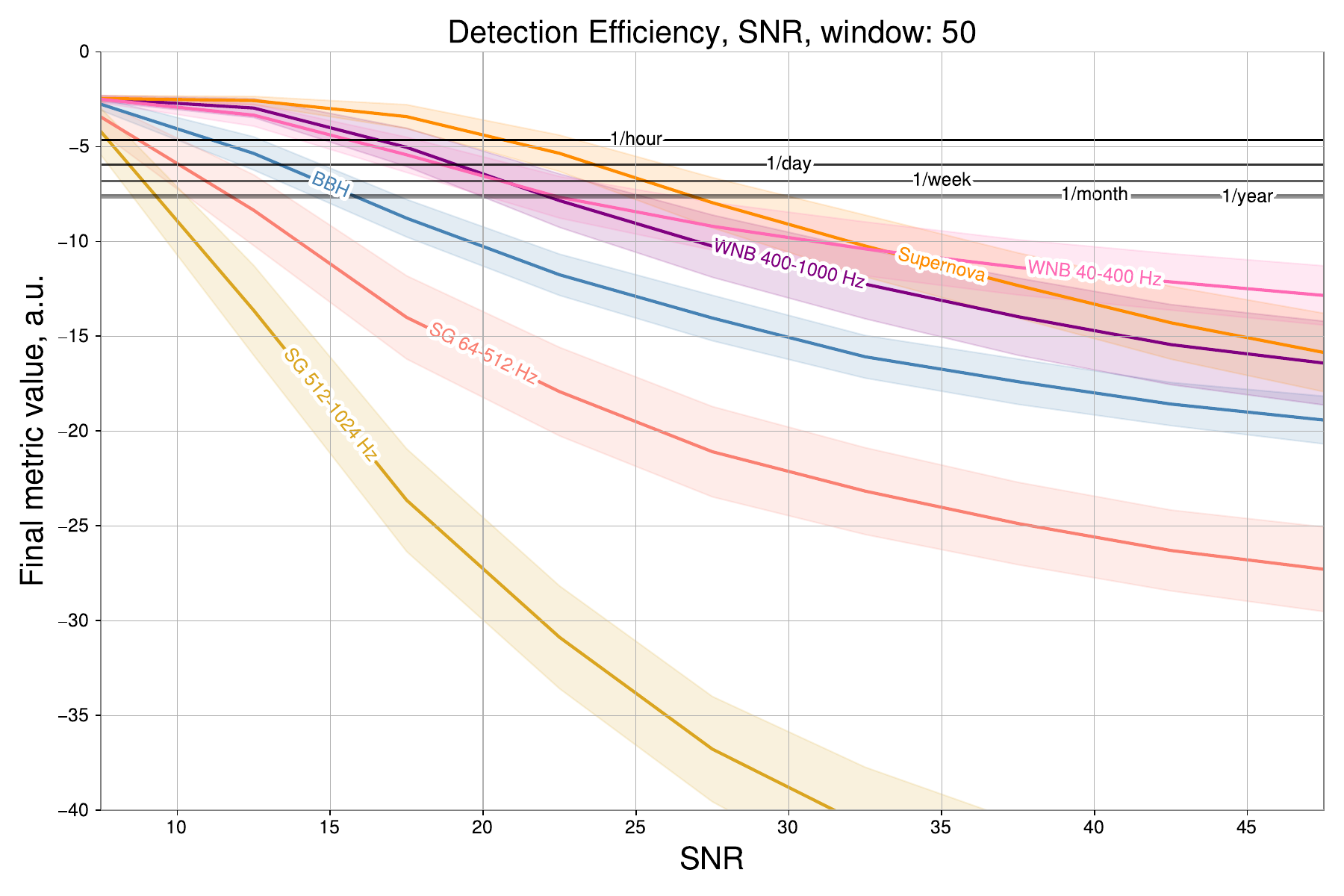}
\caption{The final metric as a function of SNR for GWAK axes training signals, BBH~(blue), SG 64--512\,Hz~(yellow), SG 512--1024\,Hz~(salmon) and for potential anomalies, WNB 40--400\,Hz~(pink), WNB 400--1000\,Hz~(purple), and Supernova~(orange). The black lines of varied width correspond to different FARs, from the FAR of 1 per hour to 1 per year. For each of the lines, the events that are below that line would be detected. As expected, the signals used to train the GWAK axes are on average detected better, eg more events with a smaller SNR are detected for a given FAR threshold.}
\label{fig:detection_efficiency}
\end{figure}

The method performance on different signals at various SNR values is shown in Fig.~\ref{fig:detection_efficiency}. For each signal type, we generate 10,000 waveforms using the SNR prior $U[5, 50]$. We then run the full GWAK algorithm and obtain the minimum metric value achieved in each injection, as shown in Fig.~\ref{fig:3_panel_plots}. We then group the injections into SNR bins of width 5, and show the average metric value as well as the $1 \sigma$ range for each bin, via the solid line and filled region respectively. The final metric values corresponding to certain false alarm thresholds are also shown. 
We first see that the three training classes - BBH, SG 64-512\,Hz, SG 512-1024\,Hz are most efficiently detected, which is expected as they have specific autoencoder models. Moving on to the anomalous signals - WNB 40-400\,Hz, WNB 400-1000\,Hz, and Supernova, we see that they are not detected as readily as the training signals, but are still able to achieve satisfactory false alarm rates around 1-2/month around 20-25 SNR.

\begin{figure}[tb]
\centering
\includegraphics[width=0.8\textwidth]{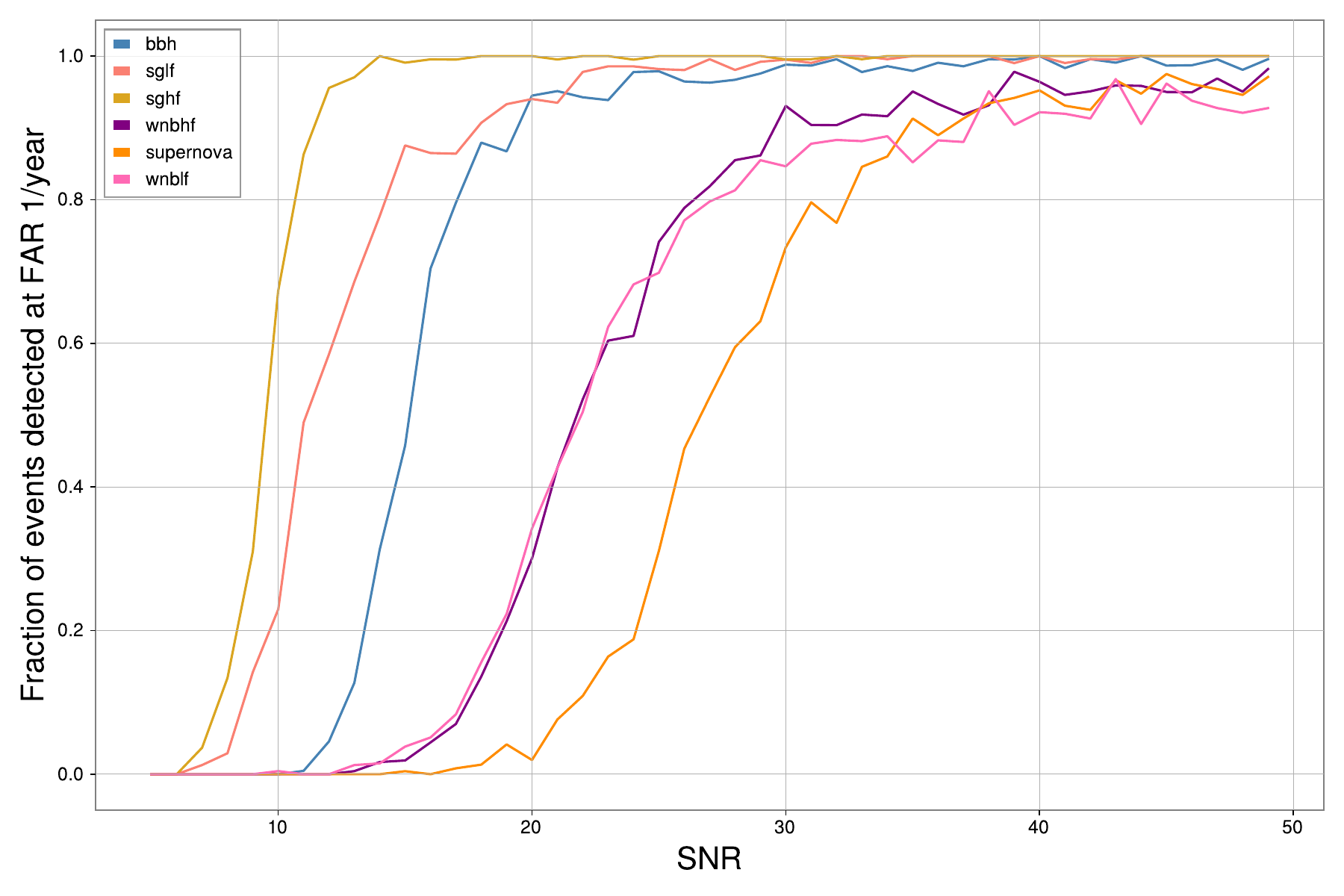}
\caption{Detection efficiency as a function of SNR for  GWAK axes training signals, and for potential anomalies. We see that the signals corresponding to training classes can be more readily detected at lower SNRs, and the anomalous signals require slightly higher SNRs to be detected.}
\label{fig:ROC}
\end{figure}

We show the detection efficiency using a receiver operating characteristic~(ROC) curve in Fig.~\ref{fig:ROC} in order to compare it with other ML techniques. 
Here, we pick a fixed false alarm rate threshold of 1/year and compute what fraction of injections, for each signal, at each SNR, have detection statistics below the threshold. Similar to Fig.~\ref{fig:detection_efficiency}, we see that the signals corresponding to training classes can be more readily detected at lower SNRs, and the anomalous signals require slightly higher SNRs to be detected. 
This graph can be compared to the one presented by MLy~\cite{skliris2022realtime}, while for the BBH and SGs we achieve similar performance, the efficiencies for WNBs and Supernovae are lower in our case. 
This is expected since the MLy algorithm was trained in a supervised manner, using WNBs during the training, while we only used those signals for finding the linear coefficients of the final metric but not as the GWAK axes.

\subsection{Comparison to a supervised search}
\begin{figure}[htb]
\centering
    \includegraphics[width=0.8\textwidth]{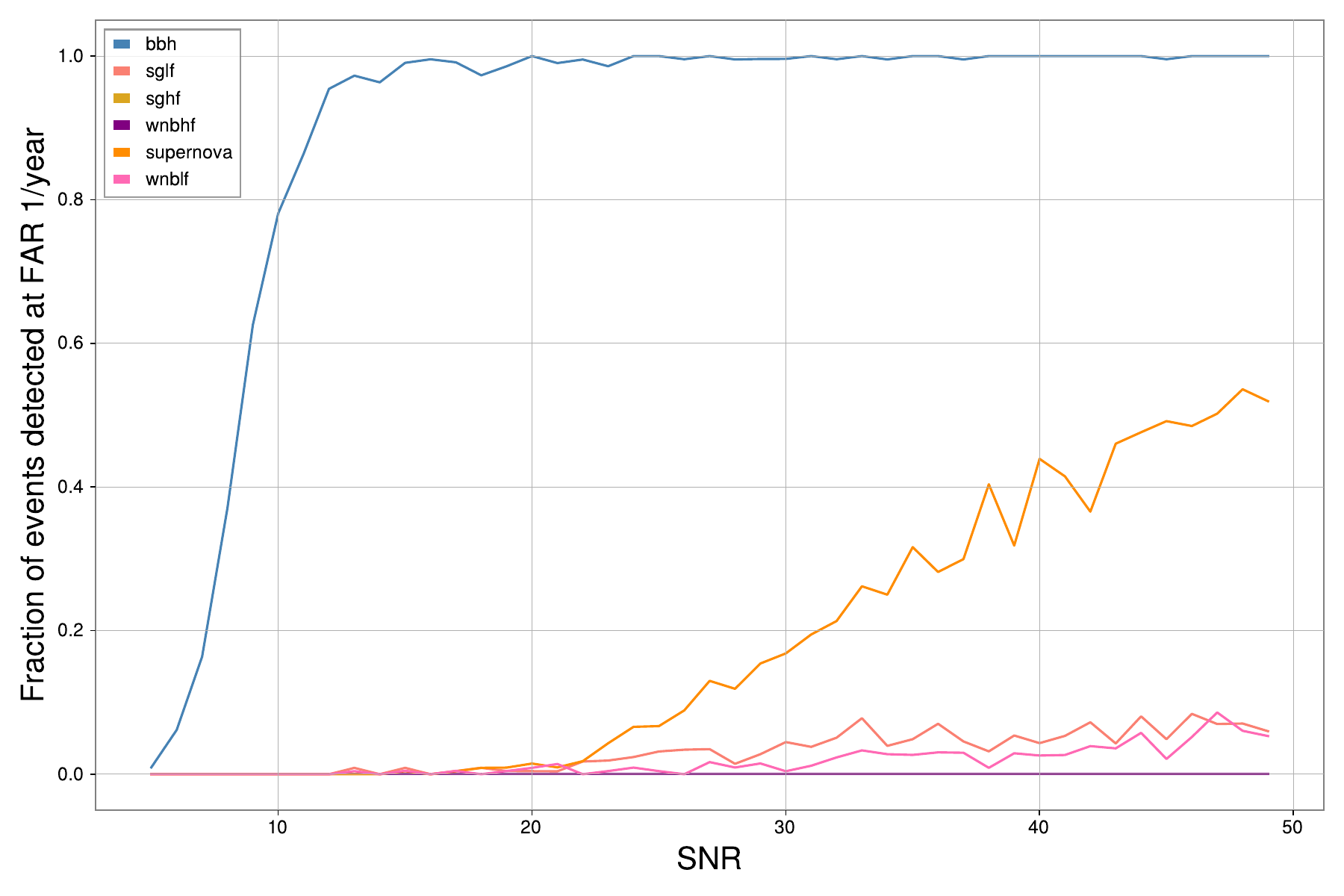}
\caption{Detection efficiency for BBH and other signals and anomalies obtained using the final metric trained in a supervised manner on BBH signals pre-processed by the GWAK algorithm. This demonstrates that the GWAK method can approach a supervised search when given this specific task.}
\label{fig:gwak_bbh_supervised}
\end{figure}

To quantify the loss of efficiency of the unsupervised GWAK method in comparison to a supervised search, we perform the following study. 
Firstly, we use pre-trained GWAK axes for the BBH search by using a small, dense network instead of a simple linear combination for the final metric. We train this network in a supervised manner, using BBH as the signal class and timeseries as the background class. 
In this way, we can quantify how much signal efficiency is lost when using a general linear combination instead of overfitting on a specific signal. 
The results are shown in Fig.~\ref{fig:gwak_bbh_supervised}. We observe that BBH detection efficiency surpasses that achieved with the linear final metric. However, as anticipated, the detection efficiency for all other signals significantly decreases. We omitted the use of a smoothing window, as it was determined to be the most efficient for BBH-supervised search. Thus, we must compare it to the BBH ROC without the application of a smoothing window, as depicted in the Fig.~\ref{fig:smoothing}.

We demonstrate that, in general, enhancing the detection efficiency for a specific signal through supervised training is feasible. However, this improvement often incurs a noticeable decline in performance for other signals. Given our objective to develop an algorithm capable of detecting unknown signals, we lack the necessary information to train it in a supervised manner. Nevertheless, in follow-up works, we may consider exploring the adoption of a more sophisticated final metric function, though we must exercise caution to prevent overfitting to the signals employed in optimizing this metric.

\subsection{Comparison of Different Smoothing Windows}
\begin{figure}[htb]
\centering
    \includegraphics[width=0.49\textwidth]{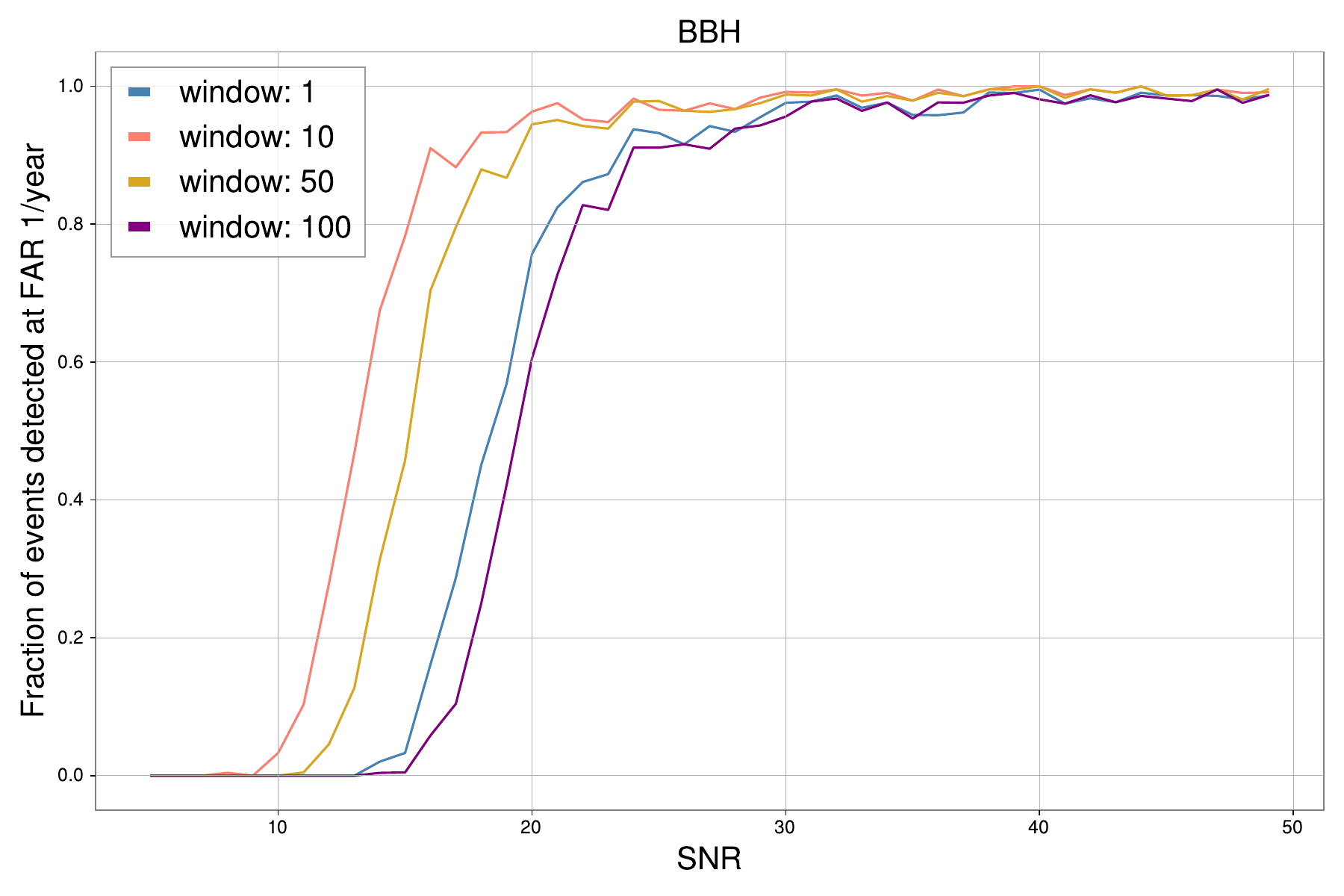}
\includegraphics[width=0.49\textwidth]{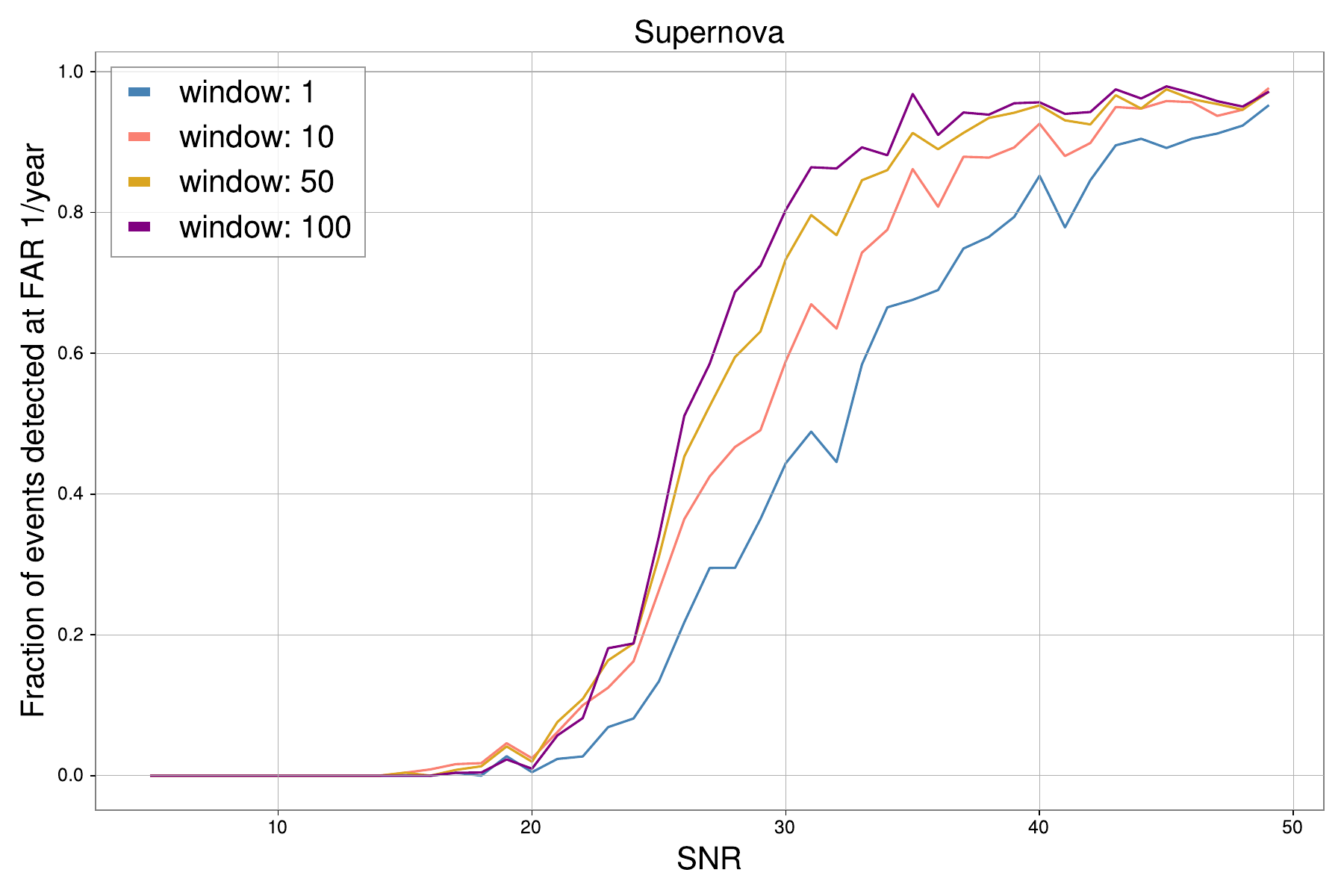} \\
\includegraphics[width=0.49\textwidth]{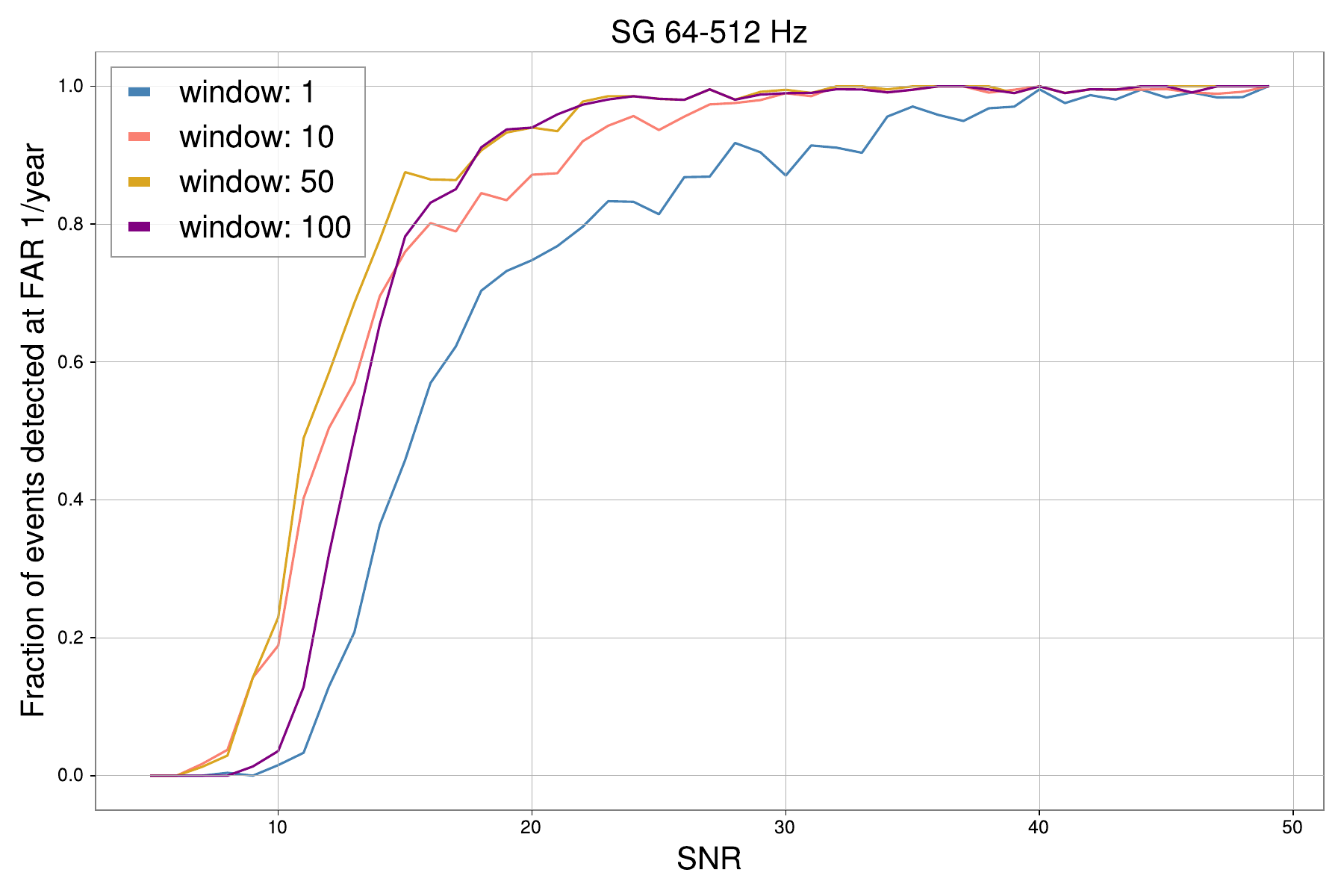}
\includegraphics[width=0.49\textwidth]{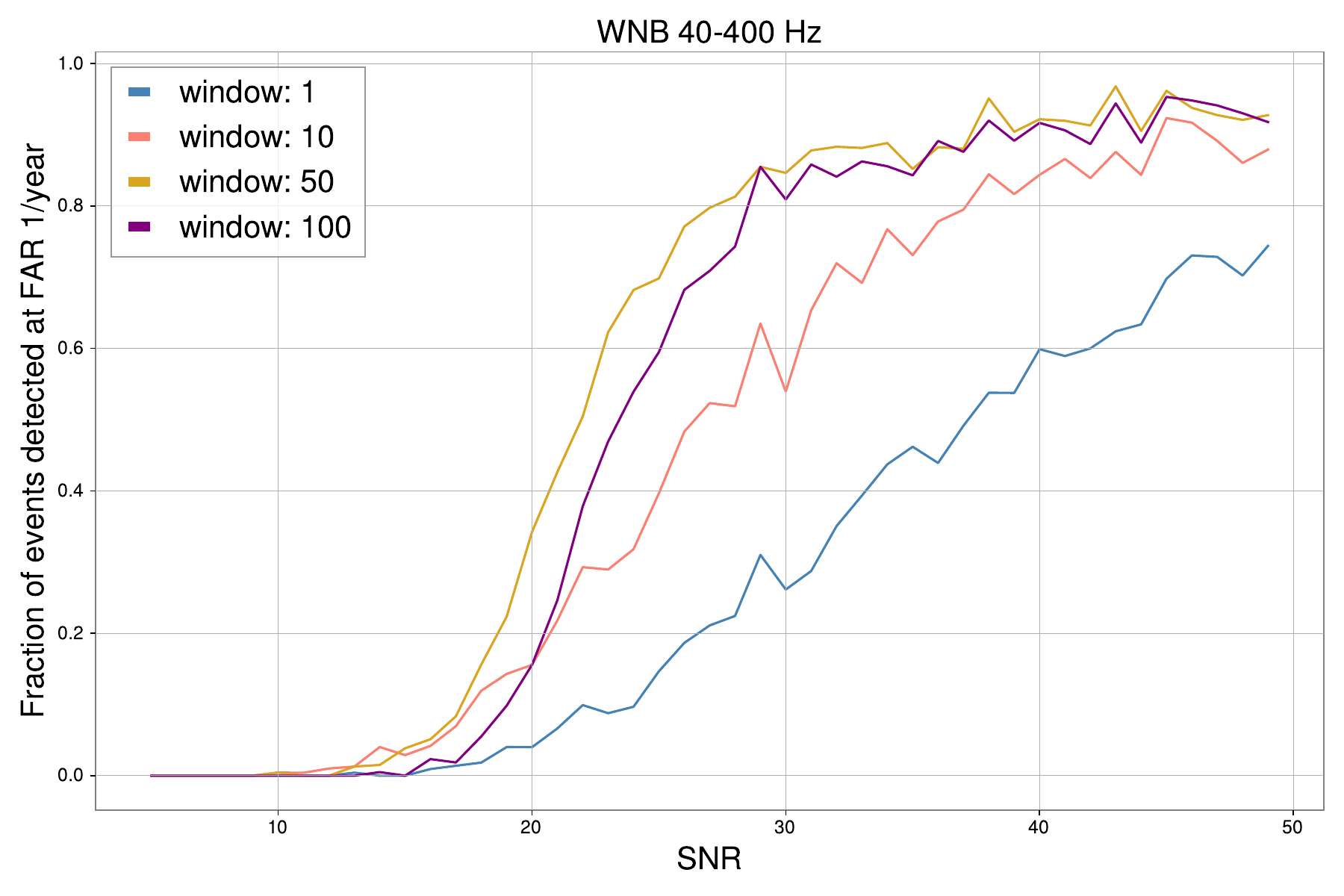} \\
\includegraphics[width=0.49\textwidth]{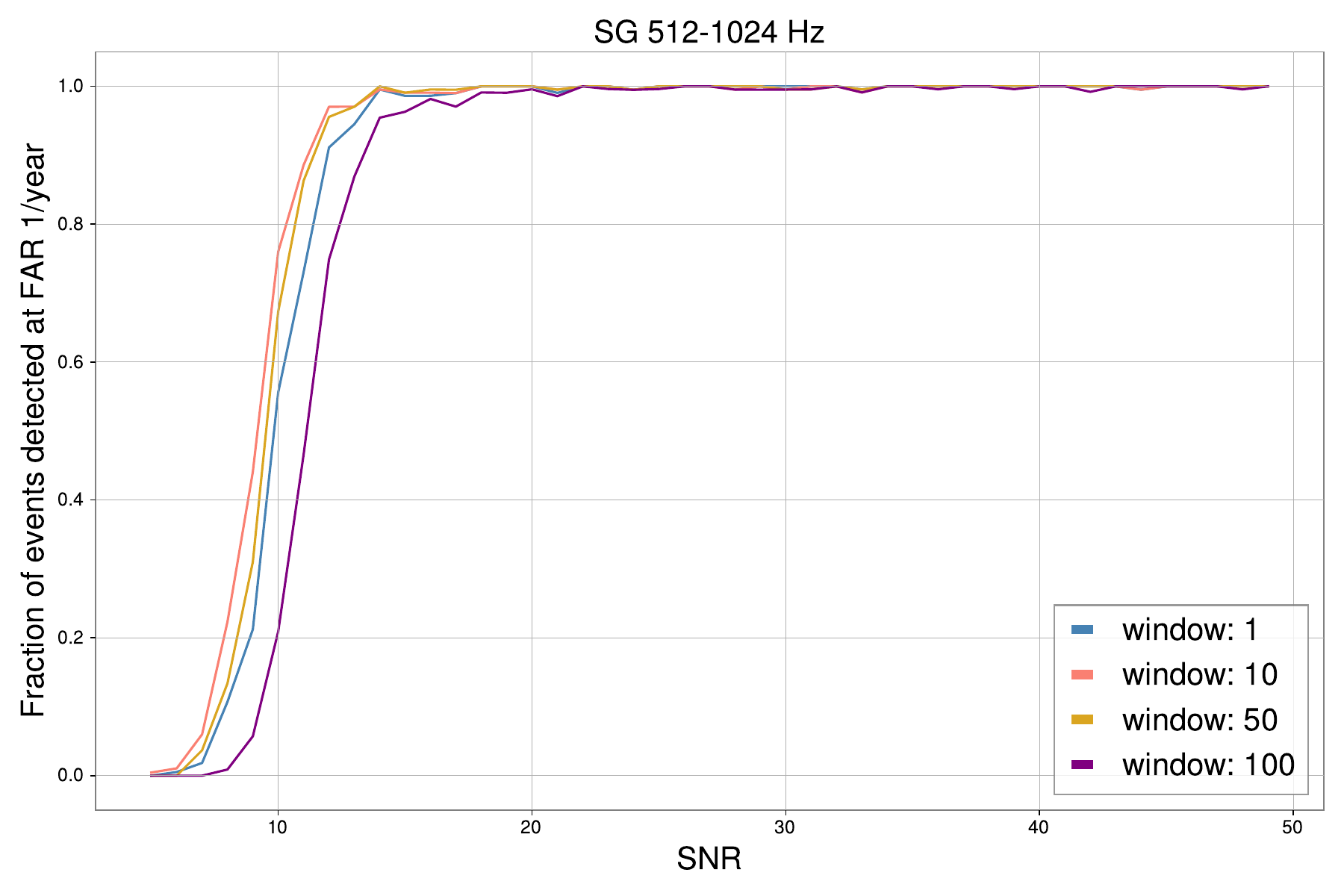}
\includegraphics[width=0.49\textwidth]{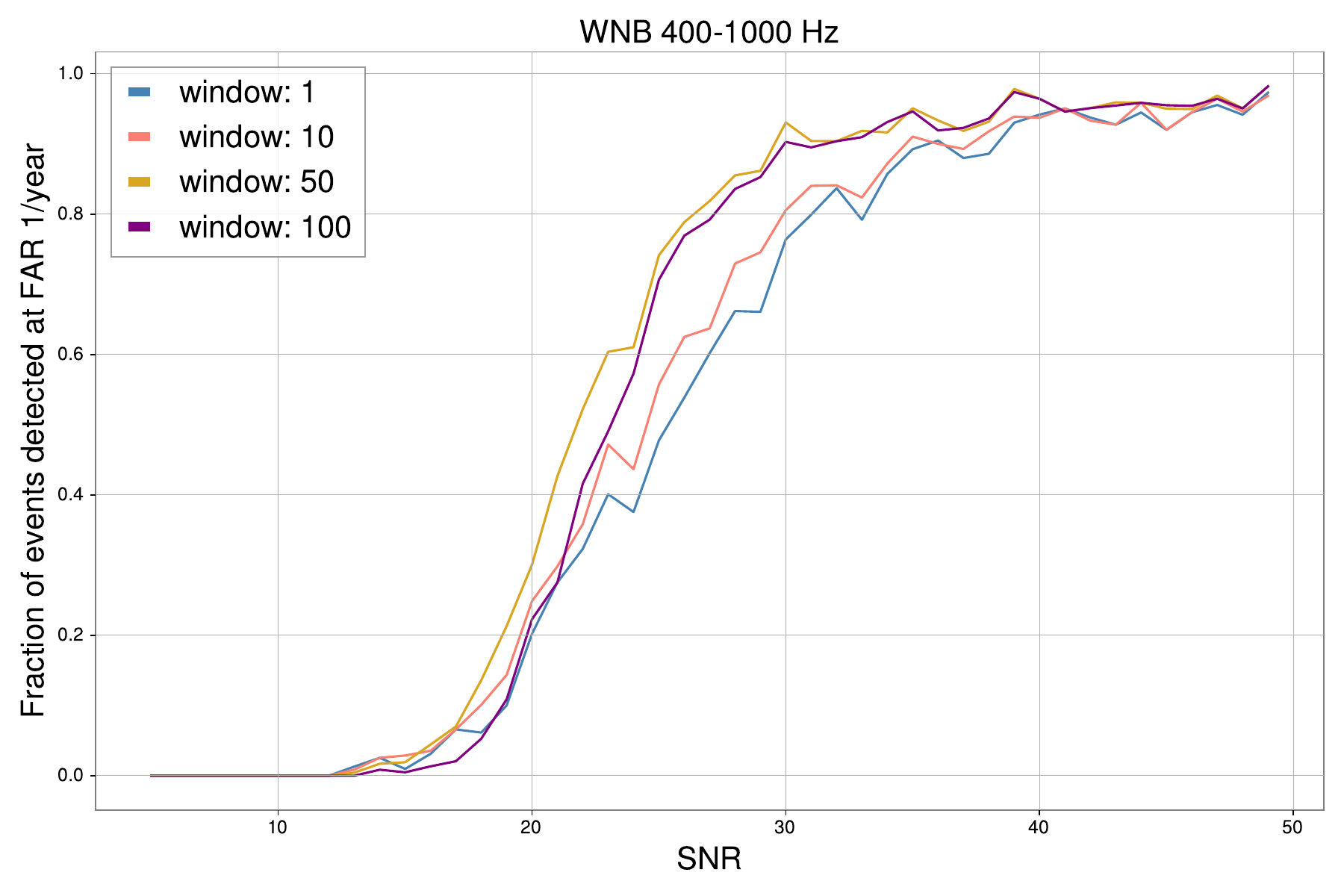} \\
\caption{Different smoothing windows for training signals~(left column) and anomalous signals~(right column).}
\label{fig:smoothing}
\end{figure}
In Fig.~\ref{fig:smoothing}, we illustrate that the optimization of detection efficiency, both for known and anomalous signals, occurs when employing smoothing kernels with sizes approximately equal to the signal length. These findings affirm that utilizing smoothing within the final metric space is an effective strategy to adapt the GWAK algorithm to diverse signal lengths

\section{Conclusions}
\label{sec:conclusions}

In this study, we utilized the Gravitational Wave Anomalous Knowledge (GWAK) method to identify anomalies in datasets acquired by ground-based GW observatories.
The GWAK method relies on the notion of introducing alternative signal priors that capture some of the salient features of new physics signatures, enabling the restoration of sensitivity even when the alternative signal is incorrect.
We separately trained five unsupervised autoencoders on a dataset consisting of normal background noise, glitches, and a collection of simulated signals that incorporate the physical characteristics of a potential new physics signature.
We then established a 12-dimensional GWAK space, comprising two reconstruction losses for each of the detector sides for each of the five autoencoders and two features representing the correlation between the detector sides and searched regions of the GWAK space for anomalous signals. 
We then combined all the 12 features into one final metric by multiplying with a corresponding coefficients from Fig.~\ref{fig:fm_weights} and summing the result. 

Our findings indicate that the GWAK method efficiently detected anomalies in the GW datasets. In particular, unmodelled sources like core-collapse supernovae and white noise bursts.
Additionally, the GWAK method could differentiate signal-like anomalies from anomalous events, such as those resulting from detector glitches.

Our proposed method demonstrates promising results, detecting these sources with high accuracy and without prior knowledge of their characteristics. These findings underscore the potential value of our approach in detecting new and unexpected sources of GW signals, while simultaneously reducing the dependence on labeled training data. In addition to serving as an unsupervised search, the machine-learning based approach allows for the implementation of the GWAK algorithm as a low-latency search tool. By detecting anomalies rapidly, alerts can be sent to electromagnetic telescopes for follow-up. 

In future work, there are many potential avenues to explore. One is using the recreation as a denoising tool instead of just a detection statistic, allowing for rapid parameter estimation. This is especially important with electromagnetic follow-up, as the telescopes need information on the source location for detailed observation. On the detection side of things, an idea is to use normalizing flows to learn the high dimensional manifolds on which signals lie and use the probability value as a score alternate to the less intuitive engineered autoencoder frequency domain features. There is also potential to improve the search by changing the way that the 12 GWAK features are reduced to a single value. While the linear fit served as a simple and explainable method and did not really possess the potential to overfit the known signals and therefore decrease sensitivity to anomalies, it suffers the drawback that potentially useful patterns between features are not exploited. A simple example is with the features $|\widetilde{H_O} \cdot \widetilde{H_R} |$ and $|\widetilde{L_O} \cdot \widetilde{L_R} |$ for a single autoencoder. For a glitch event, you would expect only one of these values to increase, corresponding to the detector in which the glitch occurred, so you would expect an asymmetry between these features for a non-astrophysical event. The opposite is true for a BBH event, for example, as the fact that it is present in both detector channels means that there should be symmetry in the BBH autoencoder features. While this is just one intuitive example, more complex relations could certainly exist. Yet another area to explore is modifying the network architecture to allow for a longer signal length to generalize the algorithm to various signal durations.

To conclude, the GWAK method displays potential as a powerful tool for detecting anomalies in GW datasets and has the potential to enhance the performance of GW anomaly detection systems.

\clearpage
\section*{Acknowledgments} 
The authors would like to thank Tino Tibaldo for the 3D figure/representation ideas. The authors would like to thank William Patrick McCormack and Jeffery Krupa for the QUAK discussions.
The authors acknowledge support from the National Science Foundation with grant numbers OAC-2117997 and CSSI-1931469.
This research was undertaken with the support of the LIGO computational clusters.
MWC and SM also acknowledge support from the National Science Foundation with grant number PHY-2010970. EM acknowledges support from the National Science Foundation with grant number GRFP-2141064. This material is based upon work supported by NSF's LIGO Laboratory which is a major facility fully funded by the National Science Foundation.

\section{Appendix}
\subsection{Backgrounds training curve}

\begin{figure}[hb]
\centering
\includegraphics[width=0.9\textwidth]{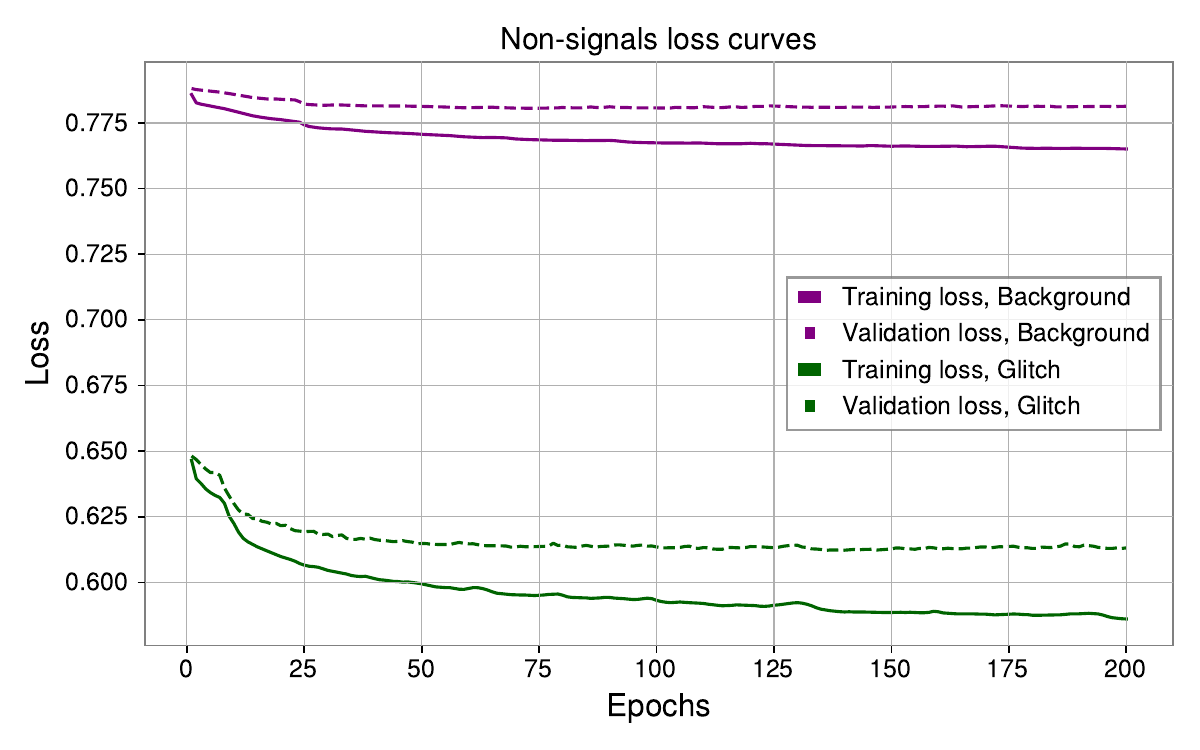}
\caption{Autoencoder training and validation losses for background classes. The training/validation losses are in bold/dashed for background~(in purple) and for glitches~(in green).}
\label{fig:background_signal_loss}
\end{figure}

For completeness, in Fig.~\ref{fig:background_signal_loss} we show the training and validation losses for the background and glitch autoencoders.

\subsection{Recreations}
\begin{figure}[htb]
\centering
\includegraphics[width=0.9\textwidth]{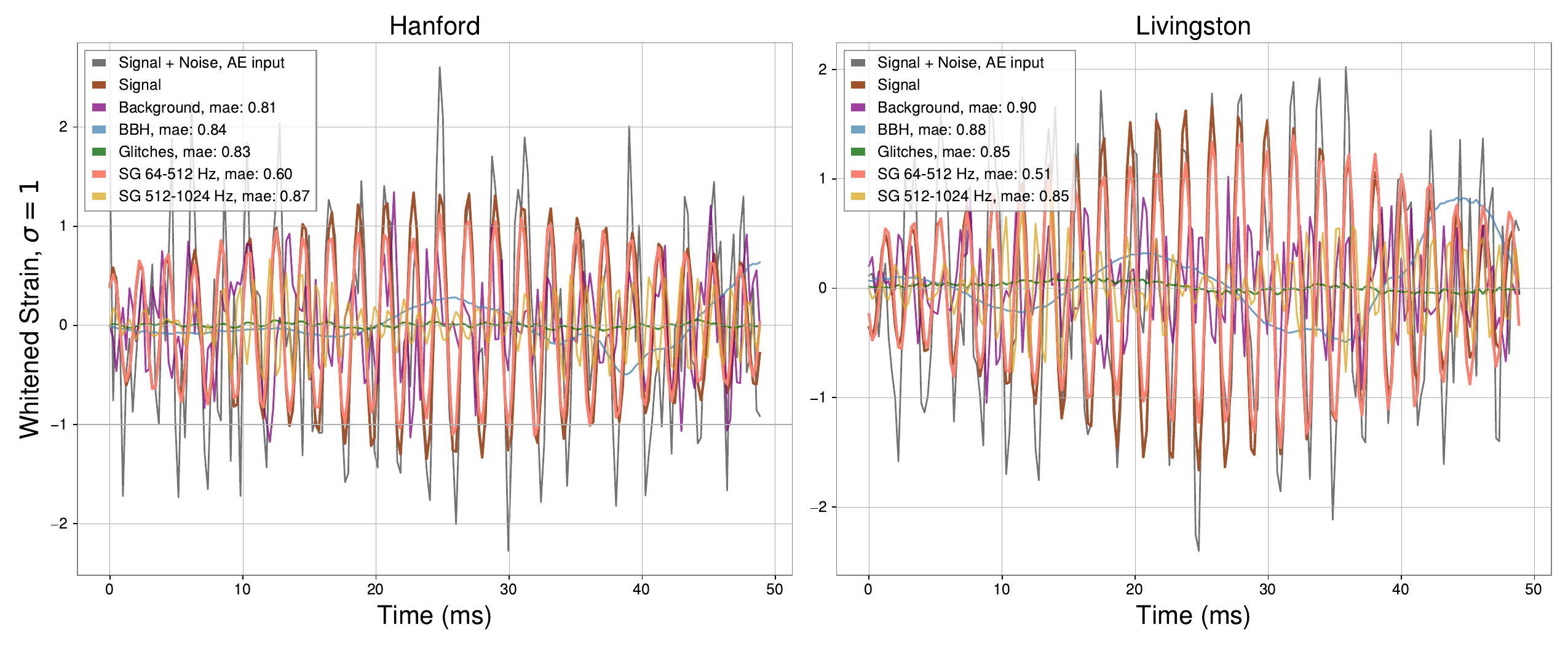}
\caption{Example of recreation plots obtained with all five pre-trained autoencoders on low frequency sine-gaussian dataset.}
\label{fig:recreation_sglf}
\end{figure}

\begin{figure}[htb]
\centering
\includegraphics[width=0.9\textwidth]{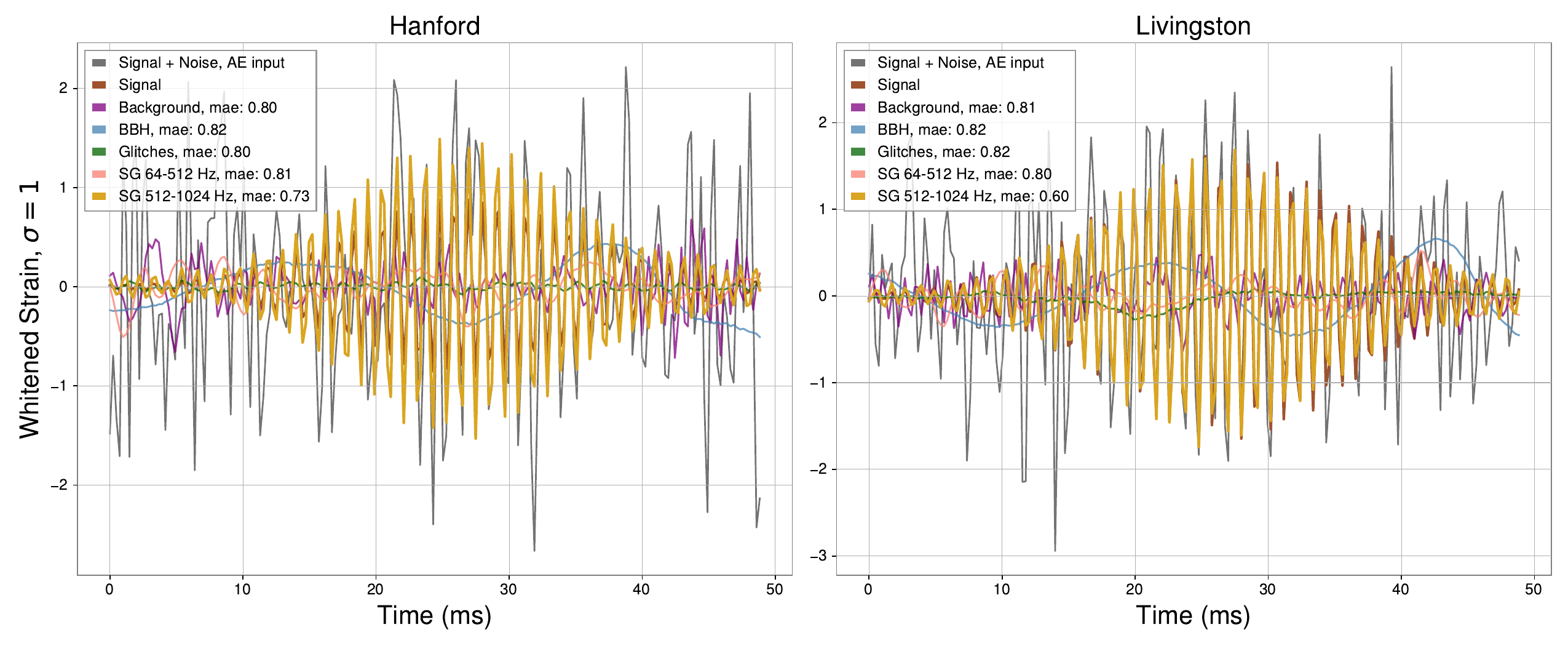}
\caption{Example of recreation plots obtained with all five pre-trained autoencoders on high frequency sine gaussian dataset.}
\label{fig:recreation_sghf}
\end{figure}

\begin{figure}[htb]
\centering
\includegraphics[width=0.9\textwidth]{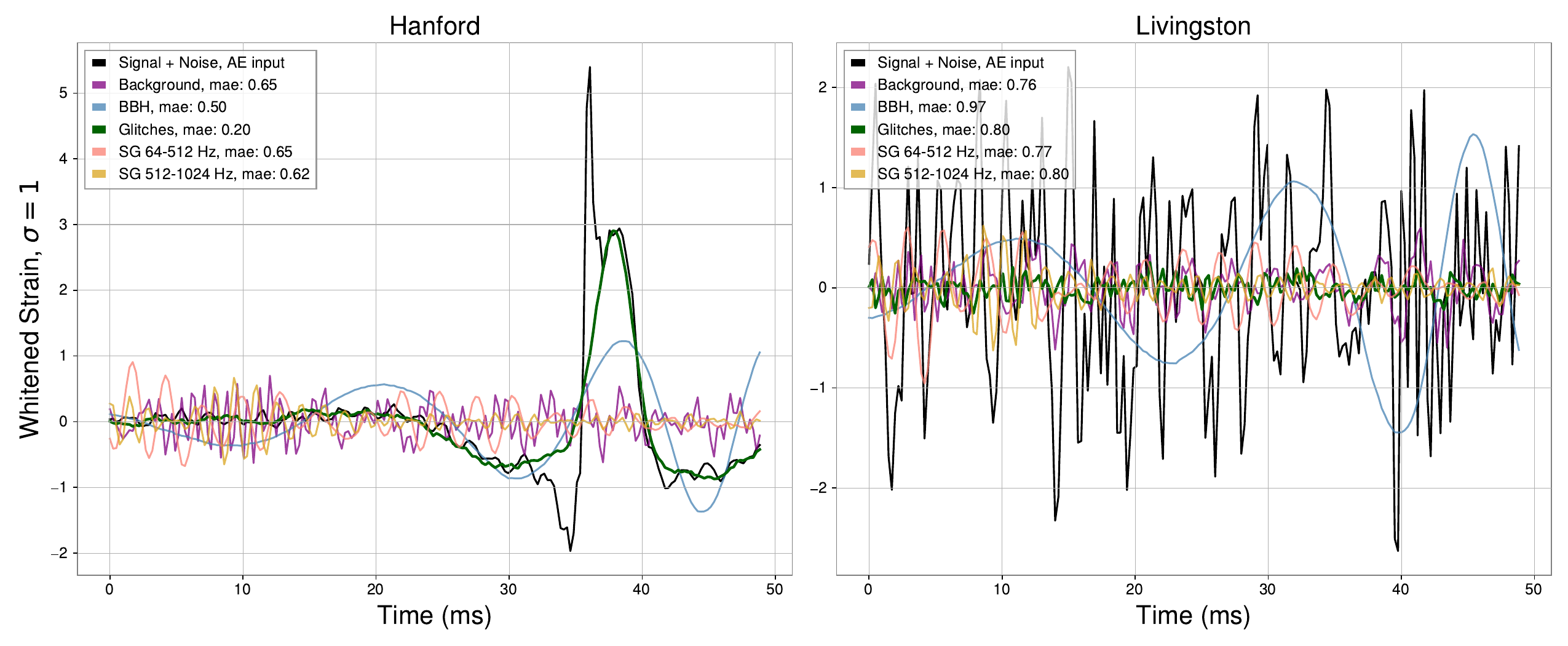}
\caption{Example of recreation plots obtained with all five pre-trained autoencoders on glitch dataset.}
\label{fig:recreation_glitches}
\end{figure}

\begin{figure}[htb]
\centering
\includegraphics[width=0.9\textwidth]{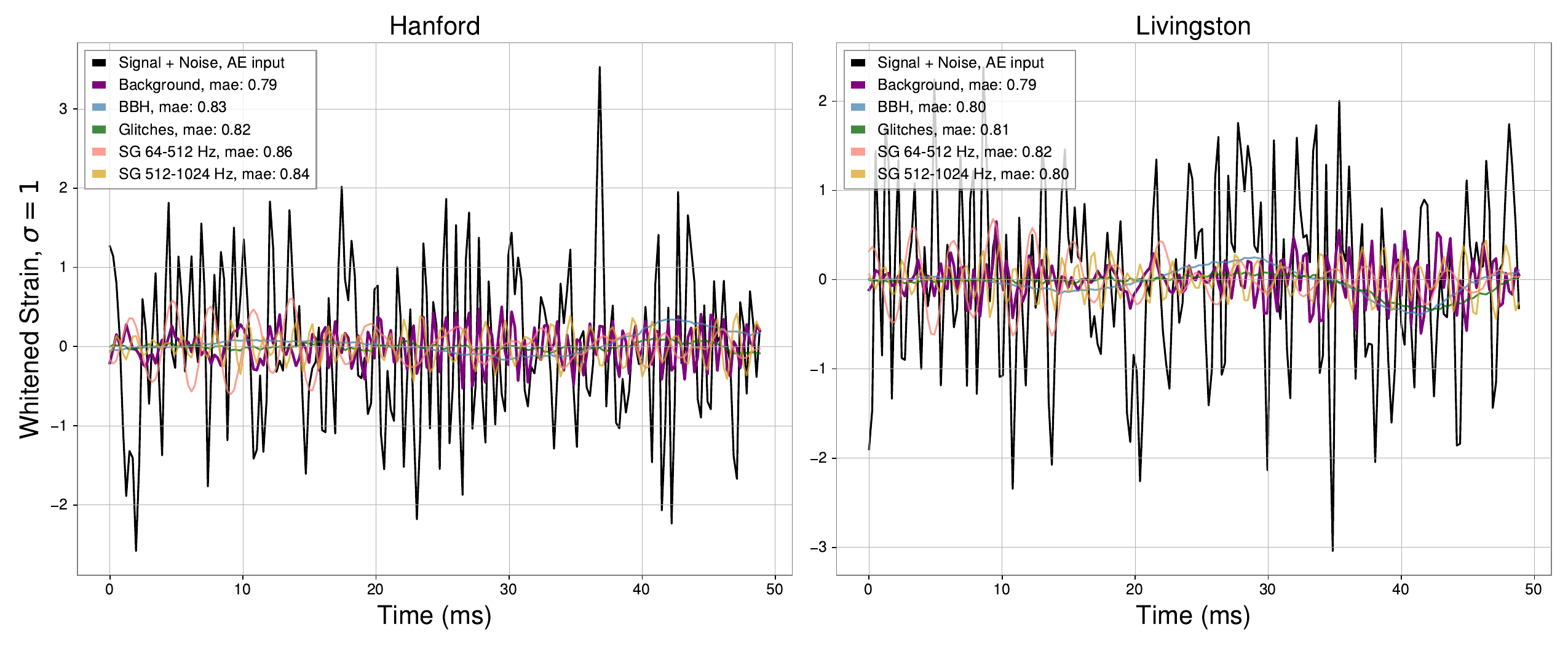}
\caption{Example of recreation plots obtained with all five pre-trained autoencoders on background dataset.}
\label{fig:recreation_background}
\end{figure}

For completeness, in Fig.~\ref{fig:recreation_sglf},~\ref{fig:recreation_sghf},~\ref{fig:recreation_glitches} and~\ref{fig:recreation_background} we show recreation plots for the low frequency SG, high frequency SG, glitches and background correspondingly. 
For each of the dataset types we can see that the corresponding autoencoder is the best in reconstructing the input, as expected.

\newpage
\section*{References}
\bibliographystyle{unsrt}
\bibliography{bib}

\end{document}